\documentclass[aps,showpacs,preprint,epsf,epsfig, nofootinbib,titlepage]{revtex4-1}
\usepackage{graphicx}
\usepackage{dcolumn}
\usepackage{bm}
\usepackage{hyperref}
\usepackage{epsfig,graphicx,color,appendix}
\usepackage{multirow}
\usepackage{array}
\usepackage{tabularx}
\usepackage{capt-of}
\usepackage{amsmath}
\usepackage{caption}
\usepackage[countmax]{subfloat}

\usepackage{footmisc}
\renewcommand{\thefootnote}{\fnsymbol{footnote}}
\usepackage{hhline}
\begin{document}
\def\t{\times}
\def\bc{{B}_{c}^*}
\def\bcp{B_c^{*+}}
\def\bcn{B_c^{*-}}
\def\bs{B_s^{*0}}
\def\abs{ \bar{B}_{s}^{*0} }
\def\bcs{B_{c}}
\def\ten{ \t 10}
\def \p {P_{1} P_{2}}
\def \ce{CKM-enhanced}
\def \cs{CKM-suppressed}
\def \cds{CKM-doubly-suppressed}
\def \tr {Transition}
\def\bsn{B_{s}^{*}}
\def\m{\mathcal{O}}
\def\b{\mathcal{B}}
\long\def\symbolfootnote[#1]#2{\begingroup%
\def\thefootnote{\fnsymbol{footnote}}\footnote[#1]{#2}\endgroup}
\def\lsim{ {\ \lower-1.2pt\vbox{\hbox{\rlap{$<$}\lower5pt\vbox{\hbox{$\sim$}
}}}\ } }
\def\gsim{ {\ \lower-1.2pt\vbox{\hbox{\rlap{$>$}\lower5pt\vbox{\hbox{$\sim$}
}}}\ } }

\font\el=cmbx10 scaled \magstep2{\obeylines\hfill \today}

\title{ \Large Flavor Dependent Effects on Rare Weak Decays of $ \bc $ and $\bsn$ Mesons}

\author{\bf Padmapriya R$^{\dagger}$ and Rohit Dhir$^{\ddagger}$}
\affiliation{\sl Department of Physics and Nanotechnology \\ SRMIST, Chennai 603203, India}
\email{$^{\dagger}${padhuspace806@gmail.com}, $^{\ddagger}$ {dhir.rohit@gmail.com};
	}

\begin{abstract}
	We investigate the effects of flavor dependence on the rare weak decays of $\bc$ and $ \bsn $ mesons to psuedoscalar and vector mesons in the final state. We use the Bauer-Stech-Wirbel model framework to calculate the flavor dependent effects of transverse quark momentum on the form factors and consequently, the branching ratios of Cabibbo-Kobayashi-Maskawa favored and suppressed modes by employing the factorization hypothesis. We find that the flavor dependent effects significantly enhance the form factors and, consequently, the branching ratios. Some of the decay channels in $\bc$ and $\bsn$ have branching ratios $\mathcal{O}(10^{-5})$ to $\mathcal{O}(10^{-7})$, which are well within the reach of current experiments.
\end{abstract}

\maketitle

\setcounter{page}{1}

\section{INTRODUCTION}
Since the observation of $B_{c}$ meson by the CDF Collaboration at Fermilab \cite{1,2}, exceptional experimental progress has been made to measure its properties and decays \cite{3,4,5,6,7,8,9,10,11,12,13,14,15,16,17}. However, only two candidates of the $B_c ~(b,\bar{c})$ family, the ground state and its first excited state , have been observed so far \cite{14}. The $B_c$ meson system is unique in the sense that it is the only known heavy meson consisting of two heavy quarks, ($b, \bar{c}$), of different flavors. Moreover, $b\bar{c}$-state can only decay weakly for its mass being below the $BD$ threshold. The study of weak decays of doubly heavy mesons are not only important for understanding of the underlying heavy flavor dynamics at $1/m_b$, $1/m_c$ and $1/\Lambda_{QCD}$ scales, but also for testing the physics within and beyond the Standard Model (SM). 

The vector, $(J^P=1^-), ~\bc $ meson is another crucial particle which is yet to be observed by the experiments. The complexity related to the observation of $\bc$ meson can be attributed, mainly, due to the difficulties in production of $\bc$ meson. Moreover, $\bc$ meson decays radiatively to $B_c$ meson via single photon emission which makes the observation even more difficult in the untidy electromagnetic backgrounds at Large Hadron Collider (LHC).   In the mean time, several ongoing and future experiments, namely, Relativistic Heavy Ion Collider (RHIC), Large Hadron Electron Collider (LHeC), International Linear Collider (ILC), the super $Z$-factory, \textit{etc.} are expected to provide large production of $B_c^{(*)}$ mesons \cite{18,19,20,21,22,23,24,25,26}. Thus, one can contemplate the discovery of $\bc$ meson in the near future. Recent theoretical analyses based on the lattice quantum chromodynamics (LQCD) estimates hyperfine splitting between $B_c$ and $\bc$ $\sim 50$ MeV \cite{27,28} which is lesser than the mass of a pion, thus forbidding strong decays. In the present experimental scenario, with high precision, the hadronic weak decays of $\bc$ meson could be easy to identify at the current experiments. Therefore, it is essential to analyze the branching ratios of $\bc$ meson through various theoretical approaches. Similarly, the study of rare weak decays of $\bsn$ meson is also gaining interest for their importance in the new physics beyond the standard model \cite{29,30,31,32,33,34,35}. Moreover, rare nonleptonic weak decays of $\bsn$ meson are expected to have larger branching ratios as compared to $\bc$ meson decays, which could be well within the reach of current experiments.  

The  $ B_{c}$ meson decays have been studied thoroughly in the literature  using different theoretical approaches \cite{36,37,38,39,40,41,42,43,44,45,46,47,48}. These studies has shown abundant hadronic channels through which $B_c$ meson can decay to lighter mesons. Similar to $B_c$ meson decays, the $\bc$ meson can also decay via both $b$-quark or $c$-quark and through weak annihilation processes. Recently, some attempts has been made to study such decays, \textit{e.g.}, using the QCD factorization hypothesis to analyze the bottom conserving decays of $\bc$ meson \cite{49}. A similar analysis has been done to study $\bc$ meson decays for the bottom changing and charm conserving mode in QCD factorization approach and light front quark model (LFQM) \cite{50}. The branching ratios of $B_{c}^{\ast}$ ${\to}$ ${\psi}(1S,2S)P$, ${\eta}_{c}(1S,2S)P$ weak decays has also been done in perturbative QCD (pQCD) approach \cite{51}. The $\bsn$ meson decays has also been studied in QCD factorization and in factorization approach \cite{52}. Most recently, the semileptonic decays of $B^{*}$, $\bsn$ and $\bc$ mesons are studied in the Bethe-Salpeter (BS) method approach \cite{53}.

Earlier, we have successfully shown the importance of flavor dependent effects on the form factors and decays of $B_c$, $ J / \psi $ and $ \Upsilon $ mesons, involving psuedoscalar ($P$) and vector ($V$) mesons in the final state, employing the factorization scheme \cite{47,48,54}. In the present work, we extend our analysis to investigate the effects of flavor dependence on $ \bc (\bsn) \to P(V)$ form factors and decays using the modified Bauer-Stech-Wirbel (BSW) Model \cite{55,56,57} framework. We also discuss and compare the QCD inspired flavor dependent effects on the form factor and decays of $\bc$ and $\bsn$ mesons. We found that the branching ratios for $ \bc \rightarrow PP/PV $ and $ \bs \rightarrow PP/PV $ decays are significantly  enhanced to $ \m (10^{-5}) \sim \m (10^{-7})$, which are well within the reach of current experiments. In addition to this, we also report several bottom changing dominant decays with the branching ratios ranging from $ (10^{-7}) \sim  (10^{-8})$, specifically for $\bsn$, which has not been considered in other approaches \cite{49,50,51,52}.

The structure of the paper is as follows: In section II, we present the theoretical framework, and in section III, we discuss the wave functions and form factors of the modified BSW model using the flavor dependent effects and QCD inspired effects on average transverse quark momenta $ \omega $. Numerical results and discussion for $ \bc $ and $ \bs $ decays are presented in section IV. Finally, we summarize our conclusions in the last section. 
\section{Theoretical framework}
\subsection{Effective Hamiltonian}
In this work, we analyze the two-body weak nonleptonic decays of $ \bc $ and $ \bsn $ mesons. The effective weak Hamiltonian 
\cite{58,59,60} for  bottom changing mode $ (\Delta b = 1) $ can be given as follows:
\begin{enumerate}
	\item [i.]$ b \rightarrow c $ decays
	\begin{equation}
	{\cal H }_{eff} =\frac {G_{f}}{\sqrt{2}} V^*_{cb}  V_{qq^{\prime}} \big[ c_1 Q_1 + c_2 Q_2 \big],     \\
	\end{equation}
	\item [ii.]$ b \rightarrow u $ decays
	\begin{equation}
	{\cal H }_{eff} =\frac {G_{f}}{\sqrt{2}} V^*_{ub}  V_{qq^{\prime}} \big[ c_1 Q_1 + c_2 Q_2 \big], \\
	\end{equation}
\end{enumerate}
where, $ q = u, c $; $ q^{\prime} = d, s $, $ V_{ij} $ are the CKM matrix elements and $G_f $ is the Fermi coupling constant. $c_{i}$ represent the Wilson coefficients, and $Q_{i}$ are current-current operators. The explicit expressions for  $ b \rightarrow c $ decays are given below: 
\begin{align}
Q_1 = \big( \bar{b}_{\alpha} c_{\beta} \big)_{V-A} \Big\{ \big(\bar{u}_{\beta} d_{\alpha} \big)_{V-A} +\big( \bar{u}_{\beta} s_{\alpha} \big)_{V-A} + \big( \bar{c}_{\beta} d_{\alpha} \big)_{V-A} + \big( \bar{c}_{\beta} s_{\alpha} \big)_{V-A} \Big\},\\
Q_2 =  \big( \bar{b}_{\alpha} c_{\alpha} \big)_{V-A} \Big\{ \big(\bar{u}_{\beta} d_{\beta} \big)_{V-A} +\big( \bar{u}_{\beta} s_{\beta} \big)_{V-A} + \big( \bar{c}_{\beta} d_{\beta} \big)_{V-A} + \big( \bar{c}_{\beta} s_{\beta} \big)_{V-A} \Big\}, 
\end{align}
where, $ \alpha $ and $\beta $ are color indices and  $ \big(\bar{q}_{\beta} q^{\prime}_{\alpha} \big)_{V-A} \equiv \bar{q}_{\beta} \gamma_{\mu} \big(1 - \gamma_{5} \big) q^{\prime}_{\alpha} $ represents the general form of $V-A$ current. One can obtain the similar expressions for $ b \rightarrow u $ transitions by replacing $c$ quark by $ u $ quark in the above expressions.\\ 
The effective Hamiltonian for bottom conserving $\big( \Delta b = 0\big) $ and charm changing $\big( \Delta C = 1\big) $  decays can be expressed as 
\begin{enumerate}
	\item[i.] $ c \rightarrow s $ decays
	\begin{equation}
	{\cal H }_{eff} =\frac {G_{f}}{\sqrt{2}} V^*_{cs}  V_{qq^{\prime}} \big[ c_1 Q_1 + c_2 Q_2 \big], \\
	\end{equation}
	\item[ii.] $ c \rightarrow d $ decays
	\begin{equation}
	{\cal H }_{eff} =\frac {G_{f}}{\sqrt{2}} V^*_{cd}  V_{qq^{\prime}} \big[ c_1 Q_1 + c_2 Q_2 \big], \\
	\end{equation}
\end{enumerate}
such that, $ q = u $; $ q^{\prime}= \{d,s\} $ and the current-current operators are,
\begin{align}
Q_1 = \big( \bar{c}_{\alpha} s_{\beta} \big)_{V-A} \Big\{ \big(\bar{u}_{\beta} d_{\alpha} \big)_{V-A} +\big( \bar{u}_{\beta} s_{\alpha} \big)_{V-A} \Big\}, \\
Q_2 = \big( \bar{c}_{\alpha} d_{\alpha} \big)_{V-A} \Big\{ \big(\bar{u}_{\beta} d_{\beta} \big)_{V-A} +\big( \bar{u}_{\beta} s_{\beta} \big)_{V-A} \Big\}.
\end{align}
The weak Hamiltonian in factorization approach express the hadronic matrix elements of current operators for nonleptonic decays via the product of meson transition matrix elements and their decay constants. The factorization assumption though not exact, but is considered to be more reliable in heavy meson decays owing to the larger energy transfers to the final states. The present work consists of the analysis of rare weak decays of  $\bc$ and $\bsn$   mesons employing the factorization hypothesis. The dominant modes in these decays proceed mainly through the tree level diagrams and thus, are expected to be least influenced by the penguin and nonfactorizable contributions. Therefore, we neglect the penguin and nonfactorizable contributions in our formalism. The QCD coefficients $a_{i}$ are generally expressed as
\begin{align}
a_1 = c_1 + \zeta c_2, \\
a_2 = c_2 + \zeta  c_1, \nonumber
\end{align}
where $ \zeta  = \frac {1}{N_c} $, at large $ N_c $ limit $ \zeta \rightarrow 0 $ and $ N_c $ is the number of color charges. We use \cite{57,60} 
$ a_1 = 1.12, a_2 = -0.26 $ and $ a_1 = 1.26 , a_2 = -0.51 $ at bottom and charm scales, respectively.

In spectator quark model all the possible decay modes can be categorized in to three classes \cite{57,58,59,60}.\\ 
\begin{description}
	\item [\textbf{Class I}]  Contains the color-favored external W-emission diagram and the decay amplitude is proportional to coefficient $ a_1$,\\
	\item[Class II]  Determined by the color-suppressed internal W-emission diagram and the decay amplitude is proportional to the coefficient $ a_2$, \\
	\item[Class III]  Caused by the interference of both the color-favored and -suppressed W-emission processes and the decay amplitude is proportional to both the coefficients $a_1$ and  $ a_2$. \\
\end{description}
\subsection{Decay amplitudes}
\subsection*{a) \boldmath $\bc \rightarrow \p $ decays}
The decay rate formula for the two body nonleptonic $\bc \rightarrow \p $ decays in the rest frame of $ \bc $ meson is given by  \cite{61,62} 
\begin{equation}
\Gamma \big(\bc \rightarrow \p) = \frac{p_{c}^{3}} {24 \pi m_{\bc}^2 } |A \big(\bc \rightarrow \p \big)|^2,
\end{equation}
where $ p_{c}$ is the three momentum of pseudoscalar meson in the final state, 
\begin{equation}
p_{c} = \frac{1}{2 m_{\bc} }  \left\{ \left[ m_{\bc}^{2} - \big(m_{P_{1}} + m_{P_{2}} \big)^{2} \right] \left[ m_{\bc}^{2} - \big( m_{P_{1}} - m_{P_{2}} \big)^{2}\right] \right\}^{\frac{1}{2}}.
\end{equation}
$m_i$ represents the masses of respective mesons. In the factorization approach, weak decay amplitude can be calculated by the product of two hadronic currents 
which is obtained by sandwiching the effective Hamiltonian between initial and final state of the wave functions of the mesons. The decay amplitude for $\bc \rightarrow \p $ can be expressed (up to the weak scale factor of $ \frac{G_{F} }{\sqrt{2} } \times$  CKM elements  $\times$ QCD factor) in terms of the reduced matrix elements \cite{57,60}: 
\begin{equation}
A \big(\bc \rightarrow \p \big) \approx \langle \p \vert {\cal H }_{eff} \vert \bc \rangle  \approx \langle P_{1} \vert J^{\mu} \vert 0 \rangle  \langle P_{2} \vert J_{\mu} \vert \bc \rangle. \\ \nonumber
\end{equation}
Here,
\begin{equation} \label{me1}
\langle P_{1} \vert J^{\mu} \vert 0 \rangle  =  -if_{p}k_{\mu}, 
\end{equation}
\begin{eqnarray} \label{me2}
\langle P \vert J_{\mu} \vert \bc \rangle &=&  \frac{1}{m_{\bc} + m_{P}} \varepsilon_{\mu \nu \rho \sigma} \varepsilon_{\bc}^{\nu} \big(P_{\bc} + P_{P} \big)^{\rho} q^{\sigma} V\big(q^2) 
-i (m_{\bc} + m_{P}) \varepsilon_{\bc}^{\mu} A_{1}\big(q^2\big) \nonumber \\
& & - i \frac{\varepsilon_{\bc}}{m_{\bc} + m_{P}} \big(P_{\bc} + P_{P} \big)^{\mu} A_{2} \big(q^2\big) 
+  i\frac{\varepsilon_{\bc} . q}{q^{2}} \big(2m_{\bc}\big) q^{\mu} A_{3} \big(q^2\big) \nonumber \\
& & - i \frac{\varepsilon_{\bc} . q}{q^{2}} \big(2m_{\bc}\big) q^{\mu} A_{0} \big(q^2\big).
\end{eqnarray}
 $ f_{p} $ and $ k_{\mu} $ are the decay constant and four momentum of the final pseudoscalar meson, respectively. $ P_{\bc} $ and $ P_{P} $ are the four momenta of the initial vector and the final pseudo scalar mesons, respectively. $ \varepsilon_{\bc}^{\mu} $ is the polarization vector of the $\bc$ meson and $ q^{\sigma} =  \big(P_{\bc} - P_{p} \big)^{\sigma} $ is the four momentum transfer.\\ 
 
From \eqref{me1} and \eqref{me2}, one can obtain the decay amplitudes for several decay modes, \textit{e.g.},
\begin{enumerate}
	\item[i.] Color enhanced mode
	\begin{eqnarray}
	A \big(\bc \rightarrow B_{s}^{0} \pi^{+} \big) 
	= \frac{G_{f} V_{sc} V_{ud}}{\sqrt{2}} ( 2 A_0(m_{\pi}^2) a_{1}^c f_{\pi^+} m_{\bc}) 
	\end{eqnarray}
	\item[ii.] Color suppressed mode
	\begin{eqnarray}
	A \big(\bc \rightarrow B^{+} \pi^{0} \big) = - \frac{G_{f} V_{dc} V_{ud}}{\sqrt{2}} (\sqrt{2} A_{0}(m_{\pi}^2) a_{2}^c f_{\pi^0} m_{\bc} ) 
	\end{eqnarray}
	\item[iii.] Color enhanced and suppressed mode
	\begin{eqnarray}
	A \big(\bc \rightarrow D_{s}^-\bar{D}^{0} \big) = \frac{G_{f}V_{sc} V_{ub}}{\sqrt{2}} \big[\big(2 A_{0}(m_{D_{S}}^{2}) a_{1}^b f_{D_{s}^{-}}  m_{\bc}\big) + \big(2 A_{0}(m_{D_{0}}^{2}) a_{2}^b f_{\bar{D}^0} m_{\bc} \big) \big]  
	\end{eqnarray}
\end{enumerate}
In above expressions, $ a_{i}^ {b} $ and $ a_{i}^{c} $ denote the QCD coefficients at bottom and charm mass scales, respectively.
\subsection*{b) \boldmath $\bc \rightarrow PV $ decays }
The decay rate formula for  $\bc \rightarrow PV $ is given by \cite{63} 
\begin{equation}
\Gamma \big(\bc \rightarrow PV) = \frac{p_{c}^{3}} {24 \pi m_{\bc}^2 } |A \big(\bc \rightarrow PV\big)|^2.
\end{equation}
The corresponding decay amplitude (up to the weak scale factor of $ \frac{G_{F} }{\sqrt{2} } \times$  CKM elements  $\times$ QCD factor) can be expressed as,
\begin{eqnarray}
A \big(\bc \rightarrow PV \big) & \approx & \langle V \vert J^{\mu} \vert 0 \rangle  \langle P \vert J_{\mu} \vert \bc \rangle.\end{eqnarray}
The matrix element, 
\begin{eqnarray}
\langle V \vert J^{\mu} \vert 0 \rangle & = & \epsilon_{\mu} f_{V} m_{V}, 
\end{eqnarray}
where, $ \epsilon_{\mu} $ is the polarization vector,  $ f_{V} $  and $m_V$ are the decay constant and mass of the final state vector meson, respectively. Furthermore, the weak amplitude of $ V \rightarrow PV $ decays can be expressed in terms of three Helicity components,
\begin{equation}
|A \big(\bc \rightarrow PV\big)|^2  = \left|H_{0} \right|^{2} +\left|H_{+1} \right|^{2} +\left|H_{-1} \right|^{2}, 
\end{equation}
such that
\begin{equation} 
H_{\pm 1} =a\pm c(x^{2} -1)^{1/2} ,\; \; H_{0} =-ax-b(x^{2} -1). 
\end{equation}
The coefficients $ a,~b$ and $ c $ represent the $ s,~d$ and $ p $ wave contributions, respectively, and are given by
\begin{eqnarray}
x & = & \frac{m_{\bc}^{2}-m_{P}^{2}-m_{V}^{2}}{2 m_{\bc} m_{V}}, \\
a & = & m_{V} f_{V} \big(m_{\bc} + m_{V} \big) A_1\big(m_{V}^{2}\big), \\
b & = & \frac{-2 m_{\bc} m_{V}^2 f_{V} A_2\big(m_{V}^{2}\big)}{m_{\bc} + m_{P}},\\
c & = & \frac{-2 m_{\bc} m_{V}^2 f_{V} V \big(m_{V}^{2}\big)}{m_{\bc} + m_{P}}.
\end{eqnarray}
Similarly, we can obtain the expressions for decay rate formulae and decay amplitudes for $ B_{s}^{*} $ by simply replacing the subscript $ c  \rightarrow s $ in the expressions for $\bc$.

\section{form factors}
The essential inputs to calculate the decay amplitudes are the form factors and decay constants. We employ the modified BSW model framework \cite{55,56,57}  to calculate the form factors from the overlap of initial and final state meson wave functions with the flavor dependent effects. The required form factors $ A_{0},~A_{1},~ A_{2}$ and $V$ at zero momentum transfer  $ (q^{2} = 0) $ are calculated from the following expressions \cite{55}: 
 \\
\begin{eqnarray} \label{int1}
A_{0}^{\bc P}(0) = A_{3}^{\bc P} &=& \int d^{2} \mathbf{{p_{T}}}
\int_{0}^{1} dx \big(\psi_{\bc}^{*^{1,0}}\big(\mathbf{p_{T}},x \big) \sigma_{z}^{0} \psi_{P} \big(\mathbf{p_{T}},x \big) \big),
\end{eqnarray} 
\begin{eqnarray} \label{int2}
I &=& \sqrt{2} \int d^{2}\mathbf{p_{T}}\int_{0}^{1} \frac{dx}{x} \big(\psi_{\bc}^{*^{-1,1}}\big(\mathbf{p_{T}},x \big) i \sigma_{y}^{1} \psi_{P} \big(\mathbf{p_{T}},x \big) \big),
\end{eqnarray}
\begin{eqnarray}
V(0) &=& \frac{m_{q_{1}} - m_{q_{1}} }{m_{\bc} - m_{P}}  I, \\
A_{1}(0)&=& \frac{m_{q_{1}} + m_{q_{1}} }{m_{\bc} + m_{P}}  I.\\
\end{eqnarray}
$ \psi_{\bc}(\mathbf{p_{T}},x) $ and $ \psi_{P}(\mathbf{p_{T}},x) $ are the wave functions of the initial and final mesons, respectively. $ m_{q_{1}} $ is the mass of non-spectator quark inside the meson. $ A_{2}(0) $ can be written in terms of the $ A_{0}(0) $ and $ A_{1}(0) $ as
\begin{eqnarray}
A_2(0)&=&\frac{2m_{\bc}}{m_{\bc}-m_{P}}A_0(0)-\frac{m_{\bc}+m_{P}}{m_{\bc}-m_{P}}A_1(0),
\end{eqnarray}
In the original BSW model \cite{55,56,57,64}, $q^2-$dependence of the form factors is assumed up to the nearest pole dominance, 
\begin{equation}
f\big(q^{2}\big) =  \frac{f(0)}{\big(1 - \frac{q^{2}}{m_{*}^{2}}\big)^{n}},
\end{equation}
where $ n = 1 $ for the monopole dependence and $ m^{*}$ is the pole mass. However, in order to be consistent with the heavy quark symmetry,  \cite{59} 
$ \big( m_{B}  + m_{D^{*}} \big)^2  = \big(m_{b} + m_{c} \big)^2 = m_{pole}^2 $, the form factors $A_0,~A_2$ and $V$ are suppose to have dipole dependence, \textit{i.e.} $ n = 2 $. Thus, we use the following $q^2-$dependence: 
\begin{equation}
A_{1}(q^{2})  =  \frac{A_{1}(0)}{\big(1 - \frac{q^{2}}{m_{A}^{2}}\big)}, \; \;  
A_{0}(q^{2})  =  \frac{A_{0}(0)}{\big(1 - \frac{q^{2}}{m_{P}^{2}}\big)^{2}},\\
\end{equation}
\begin{equation}
A_{2}(q^{2})  =  \frac{A_{2}(0)}{\big(1 - \frac{q^{2}}{m_{A}^{2}}\big)^{2}}, \; \; 
V(q^{2})  = \frac{V(0)}{\big(1 - \frac{q^{2}}{m_{V}^{2}}\big)^{2}}.
\end{equation}

The wave function of the meson is given in the form of solution of the scalar relativistic harmonic oscillator \cite{55},  
\begin{equation} \label{psi}
\Psi_{m} \big(\mathbf{p_{T}},x \big) = N_{m} \sqrt{x(1-x)} \exp \big(-\frac{\mathbf{p_{T}}^{2}}{2 \omega^{2}} \big) \exp \big(- \frac{m^{2}}{2 \omega^{2}} \big(x - \frac{1}{2} - \frac{m_{q_{1}^{2}} -m_{q_{2}^{2}}}{2 m^{2}} \big)^{2} \big).
\end{equation}
 $N_{m}$ is the normalization constant, $m$ is the mass of meson, $ m_{q_{1}} $ and $ m_{q_{2}} $ denotes the mass of quark and anti-quark, respectively,  and $ x = \frac{p_{z}}{\textbf{P}} $, $ \mathbf{p_{T}} = \big(p_{x}, p_{y}\big) $ are the fraction of the longitudinal and transverse quark momentum of the non-spectator quark within the meson. $\textbf{P}$ is the momentum of initial meson and $ \omega $ is the dimensional quantity that represent the average transverse quark momentum, $ \langle p_{T}^{2} \rangle $. 

In BSW approach, meson wave function depends, apart from the constituent masses, upon only one parameter, \textit{i.e.} $\omega$, the transverse momentum of quark inside the meson. The form factors (at $q^2=0$) are determined by  the wave function overlap via expressing current $J^\mu$ in terms of creation and annihilation operators. The appropriate components of current operator in \eqref{me2} are considered to express the form factors in terms of the space integrals \eqref{int1} and \eqref{int2}, respectively \cite{55,56,57}. These space integrals depend upon the precise form of the wave functions, which in turn are flavor dependent. However, in the original BSW model calculations, the transverse quark momentum inside the meson has been fixed at $ \omega= 0.40$ GeV, same for both the initial and the final state mesons. The calculated values of the form factors at $ \omega= 0.40$ GeV are shown in Tables \ref{t1} and \ref{t2}. It is well known that the meson transition form factors are sensitive to the choice of parameter $ \omega $. Since, the wave functions for initial and final state mesons are different (in flavors), the overlap integrals and hence, the form factors are expected to be flavor dependent. In the following sub-sections, we will discuss the effects of flavor dependence  on $\omega$ and consequently, on the form factors. 

\subsection{Flavor Dependent Effects on \boldmath $\bc$ and $ B_{s}^{*}$ decays}
\begin{figure}[h]
	\centering
	\includegraphics[width=0.8\textwidth]{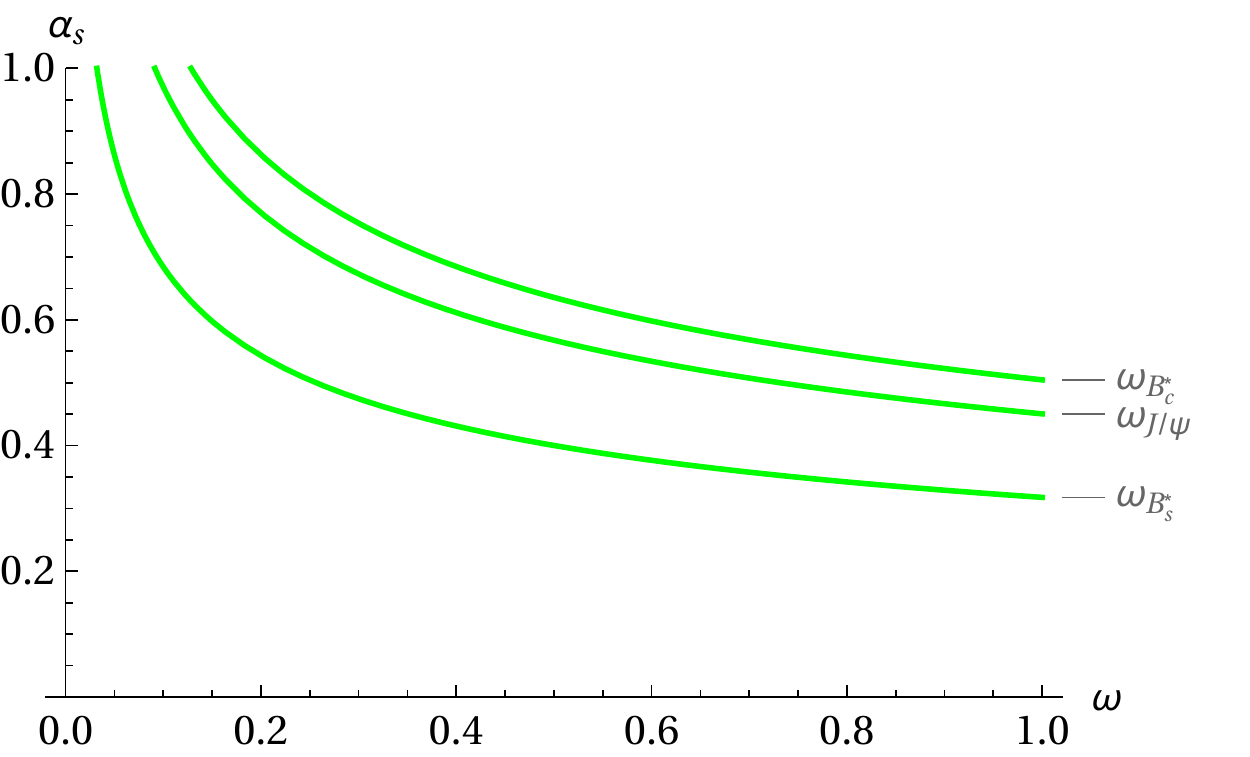}
	\caption{The variation of flavor dependent parameter $\omega$ w.r.t strong coupling constant $
	\alpha_{s}$.}
	\label{fig1}
\end{figure}

Previously, the flavor dependent effects on the parameter $ \omega $ has been studied in detail for $ \bcs \rightarrow PP, PV $ \cite{47,48} and $ J/\Psi, \Upsilon \rightarrow PP,PV $ decays  \cite{54}. 
In this section, we will extend our analysis to calculate the  form factors of $ \bc $ and $ B_{s}^{*}$  with the flavor dependent effects. Based on the dimensionality arguments, the parameter $ \omega $ can be related to the square of the ground state wave function at the origin for the corresponding meson by the following ansatz,
\begin{equation} \label{dim}
|\Psi(0)|^{2} \propto \omega^{3}.
\end{equation}
The $ |\Psi(0)|^{2} $ can be extracted from the  well known hyperfine splitting relation for the meson masses \cite{65}. 
\begin{equation} \label{hfs}
|\Psi(0)|^{2} = \frac{9 m_{i}m_{j}}{32 \alpha_{s} \pi} (m_{V} - m_{P})
\end{equation}    
where $ m_{i} $ and $ m_{j} $ are the quark masses, $ m_{V}$ and $ m_{P} $ are the masses of the vector and pseudoscalar mesons, respectively, and $ \alpha_{s} $ is the strong coupling constant. The scale or momentum ($q^2$) dependence of $\alpha_{s}$, near the asymptotic freedom, for high $q^2$ (at short distances) is well examined and precisely measured \textit{i.e.}, $\alpha_{s}(M^2_Z) = 0.1185 \pm 0.0006$ at mass scale $M_Z = 91.19$ GeV \cite{66}. However, the long-distance behavior of $\alpha_{s}$ for ($ q^2\le 1$ GeV) \textit{i.e.} infrared (IR) region is not yet well defined \cite{67}. Thus, its very difficult to determine $\alpha_{s}$ at $u,~d, ~\text{and}, s$ quark mass scale. The long distance understanding of strong coupling constant is of much importance for many QCD phenomena like hadron structure, quark confinement, hadronization processes \textit{etc} \cite{68}. In literature \cite{67,68,69,70,71}, the values for $\alpha_{s} ~(\text{at} ~ u,~d,~s ~\text{quark mass scale})$ varies from $ 0.5 \sim 0.7$ for $q^2 < 1$ GeV. However, we use   
\begin{eqnarray}
\alpha_{s} = 0.60 ~\big(\text{at} ~ u,~d,~s ~\text{quark mass scale} \big), ~
\alpha_{s}(m_{c}) = 0.31, ~\alpha_{s}(m_{b}) = 0.21;
\end{eqnarray} 
in the present calculation at different mass scales. These values have been calculated from the RunDec-Mathamatica package \cite{72} and are consistent with the values used in literature. To obtain the numerical values of the form factors, we use the following quark masses (in GeV), \begin{equation}
m_{u} = m_{d} = 0.31, \; \; m_{s} = 0.48,\\
m_{c} = 1.66, \; \; m_{b} = 5.00
\end{equation}
We wish to point out that although the strong coupling constant has been fixed at $\alpha_{s} = 0.60 ~\text{for} ~( u,~d,~s)$ lighter quarks, the  SU(3) symmetry breaking has partially been taken care of by using different masses for $u(d)$ and $s$ quarks. The calculated values of $|\Psi(0)|^{2}$  and $ \omega $ are given in the columns 2 and 3 of the Table \ref{t3}. Even though the size of $ \pi $ and $K$ meson differs, the value of  $|\Psi(0)|^{2}$ doesn't vary much since the variation could only be observed at fourth decimal place. It may be noted from \eqref{hfs} that the square of the meson wave function at the origin, $ |\Psi(0)|^{2} $, depends upon meson and constituent quark masses and strong coupling constant. Since, the QCD forces by very nature are flavor conserving and independent of flavor, the estimates of $ |\Psi(0)|^2$  from hyperfine splitting has  been successfully used to the study properties of hadrons \cite{73,74,75,76,77,78,79}. In fact, the magnitude of $ |\Psi(0)|^2$ for the $b\bar{c}$ system, in \cite{77}, is evaluated via interpolation between $c\bar{c}$ and $b\bar{b}$ systems to parametrize $ |\Psi(0)|^2$  as some power $p$ of the reduced mass. The results obtained in \cite{77,79} match well for $ |\Psi(0)|^2$ obtained in our case within $10-15\%$. Since, the experimental masses of mesons are very precisely measured, we expect that $ |\Psi(0)|^{2} $ determined from \eqref{hfs} will be quite reliable. We fix the value $ \omega(D) = 0.41 $ by using the well measured form factor $ F_{0}^{DK} = 0.75 \pm 0.02 $ \cite{66}.  
The variation of $ \omega $ (as shown in column 3 of  Table \ref{t3}) is justified based on dimensionality argument \eqref{dim}, where we incorporate the flavor dependent effects through the size of the meson. It may be noted that the value of parameter $ \omega $ increases with decreasing  meson size. In order to understand the dependence of parameter $\omega$ from hyperfine splitting for $b\bar{c}$ states with strong coupling constant, we plot $\omega$ with $\alpha_{s}(q^2)$ as shown in Fig.\ref{fig1} for $\bc, ~\bsn$ and $J/\psi$ mesons. It is interesting to note that the parameter $\omega_{\bc}$ seems to be almost independent of strong coupling constant. 

\begin{figure}[h]
	\centering
	\includegraphics[width=1.\textwidth]{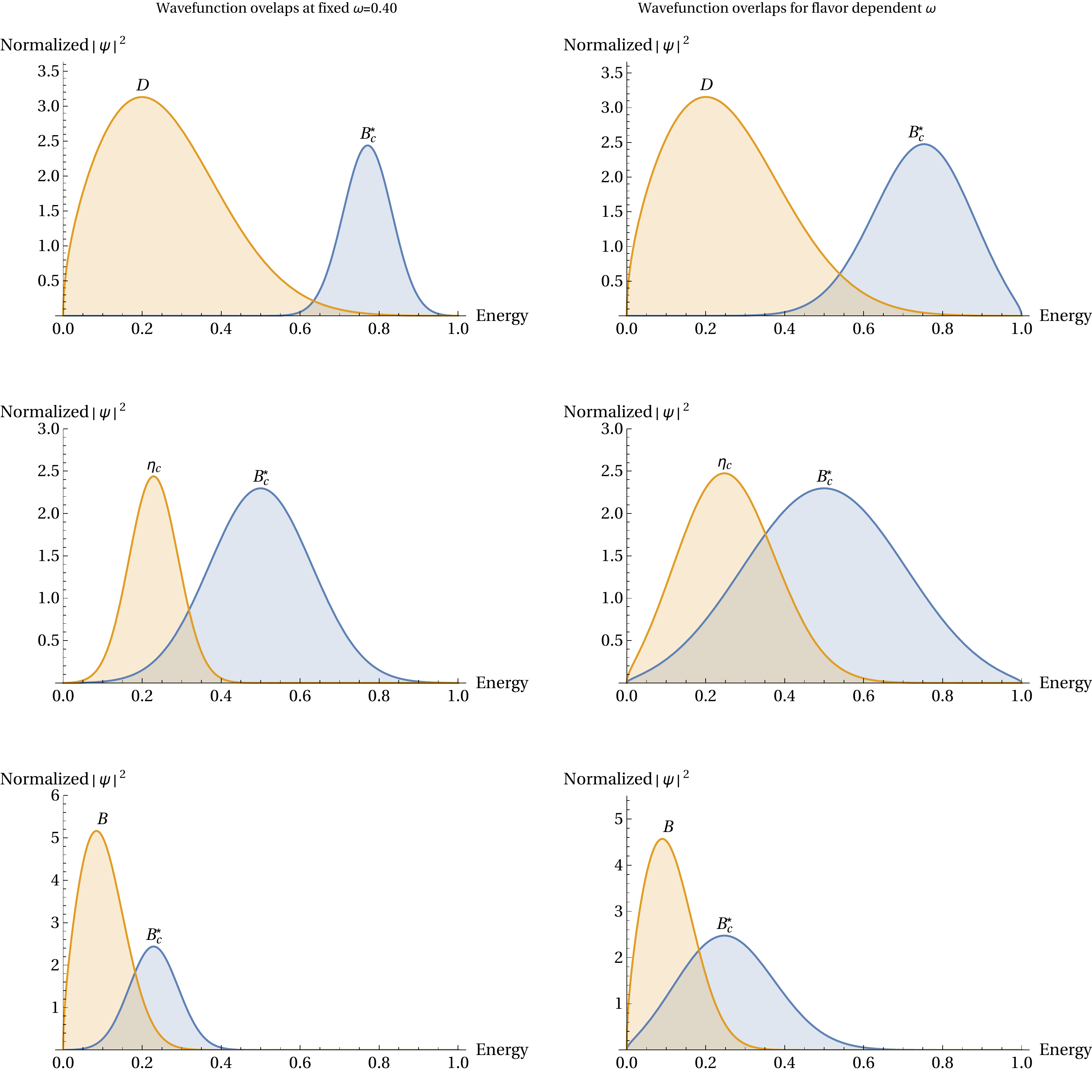}
	\caption{The wave function overlaps $ \bc $ meson with $D,~B,~ \eta_{c} $  mesons at fixed $\omega = 0.40$ and for flavor dependent $\omega$'s.}
	\label{fig2}
\end{figure}

Aforementioned, the numerical values of the various transition form factors depend upon the wave functions overlap between initial and final mesons. The meson wave function depends only on parameter $\omega$, which in fact is flavor dependent. Therefore, we use the flavor dependent $\omega$ to determine the transition form factors. The obtained results are given in the Tables \ref{t1} and \ref{t2} for the flavor dependent case. To highlight the significance of flavor dependent effects, we plot the wave function overlaps for $ \bc \rightarrow D $, $ \bc \rightarrow \eta_{c} $ and $ \bc \rightarrow B $, transitions as shown in Fig.\ref{fig2}. It can be seen that the effective wave function overlaps are significantly enhanced for flavor dependent $\omega$ as compared to fixed $\omega=0.40$ GeV. Since, the form factors are evaluated from the space integrals of wave function overlaps of initial and final mesons, the numerical values of the corresponding form factors has increased significantly (as shown in Tables \ref{t1} and \ref{t2}) in comparison to flavor independent case.

\subsection{QCD Inspired calculation of \boldmath$ \omega $}
\begin{figure}[h]
	\centering
	\includegraphics[width=0.8\textwidth]{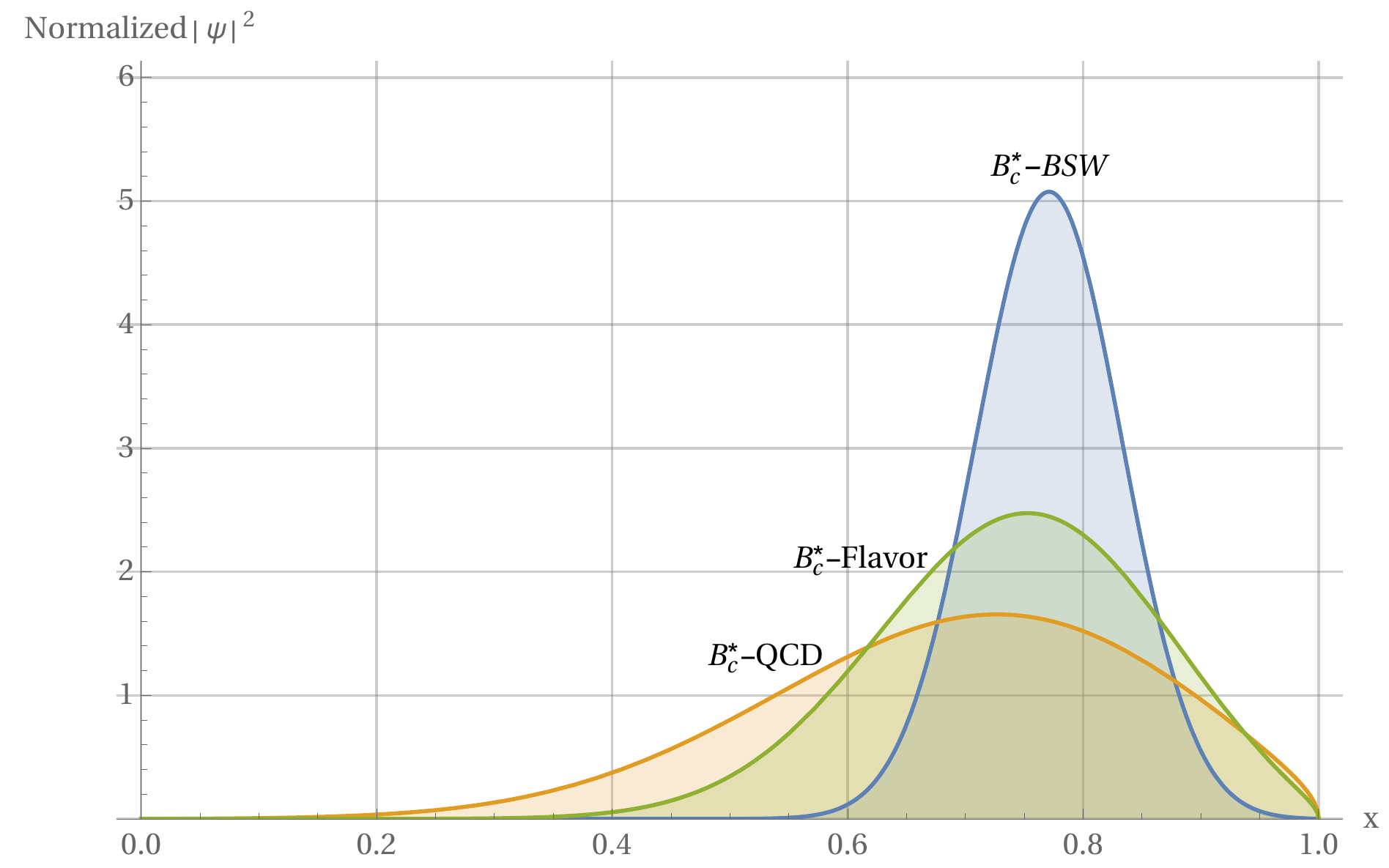}
	\caption{The distribution amplitude of $ \bc $ meson for BSW (fixed $\omega=0.40$ Gev), flavor dependent and QCD inspired cases.}
	\label{fig3}
\end{figure}

In the heavy quark limit, the distribution amplitude of $\bc$ meson shows a peak near $x \simeq x_0 = m_b/m_{B_c}$ (as shown in Fig. \ref{fig3}). Thus, the width of the peak decreases as the mass of heavy quark $m_Q$ becomes larger. The average transverse quark momentum (velocity) equals to that of the heavy meson such that the average $x \to 1$, up to corrections of order $p_T/M_{B_c} $. It is easy to interpret since $\langle p_T^2 \rangle \approx \Lambda_{QCD}^2$, the wave function $\psi(x, p_T )$ vanishes if $\langle p_T^2 \rangle \gg \Lambda_{QCD}^2$ with peak at as $x \to 1$.  Therefore, the net effect is that the transition matrix element is significant only in the region near $x \to 1$. In the light of above discussion, looking at the typical inverse size of the system ($\Lambda_{QCD}$) under scrutiny, the average transverse quark momentum for a system of two heavy quarks, including $b \bar{c}$, can be given by
\begin{equation} \label{qcdw}
\omega_{\Lambda} \approx \alpha_{s}M,
\end{equation}
where \textit{M} is the mass of the heavy meson. It may be noted that in non-relativistic QCD approximations \cite{80}, for quarkonium like states,  a potential constructed from a one-gluon exchange is expected to be sufficient. Such states are considered as a non-relativistic system bound via a Coulomb potential, whose (inverse) size is of the $m_Q \alpha_{s}\sim M_Q|v|$, since the non relativistic velocity of the quarks in this system $|v| \sim \alpha_{s}$ \cite{81,82}. However, the relativistic corrections of $\mathcal{O}(v^2)$ are ignored in such approximations, which could lead to uncertainties as large as $30\%$\cite{83,84}. Furthermore, for the system of heavy quarks, the coupling constant $\alpha_{s} (m_Q)$ is usually small, which implies the length scale is comparable to the Compton wavelength $\sim 1/m_Q$, at such scales the strong interactions are perturbative. However, hadronization as a matter of course is non-perturbative, therefore, the calculation of the hadronic matrix elements is nontrivial due to the non-perturbative effects \cite{80,81,82}. Besides the above arguments, using \eqref{qcdw} we calculate $ \omega_{\Lambda} $ for various mesons as shown in column 4 of Table \ref{t3} and thereupon, we obtain the form factors for the QCD inspired case as given in the Tables \ref{t1} and \ref{t2}. Here again, we take $ \alpha_{s} $ for $b\bar{c}$-system at bottom mass scale. Interestingly, with $ \alpha_{s} = 0.21 $ for $ \bc({B_{c}})$ system yield $ \omega_{\Lambda} = 1.33 $, which is substantially large when compared with flavor dependent case. It may be noted that the both flavor dependent and QCD inspired calculation points to the fact that $ \omega$ could be reasonably large as compared to  $ \omega = 0.40 $ GeV for heavy mesons.  We plot the distribution amplitude for $\bc$ wave function for  all the three cases of $\omega$, as shown in Fig. \ref{fig3}. The width of the wave function increases with increasing value of parameter $\omega$ with peak shift towards the lower $x$ implying $\langle p_T^2 \rangle \approx \Lambda_{QCD}^2$. Therefore, the flavor dependent effects in case of QCD inspired $\omega_{\Lambda}$ and the fixed $\omega=0.40$ GeV can be treated as limiting cases for $b\bar{c}$-system with average close to flavor dependent $\omega=0.85$ GeV.
 
\section{Numerical Results and discussion}
To obtain the numerical results, we have used the $ SU(3)$ pseudoscalar and vector nonets for light meson sectors with the following mixing scheme between singlet and octet states: \\For the pseudoscalar mesons, 
\begin{eqnarray}
\eta^{\prime} & = & \frac{1}{\sqrt{2}} \big(u \bar{u} + d \bar{d}\big) ~ cos\phi_{P} - s \bar{s} ~ sin\phi_{P}, \\
\eta & = & \frac{1}{\sqrt{2}} \big(u \bar{u} + d \bar{d}\big) ~ sin \phi_{P} - s \bar{s} ~ cos\phi_{P}.
\end{eqnarray} 
and for the vector mesons,
\begin{eqnarray}
\phi & = & \frac{1}{\sqrt{2}} \big(u \bar{u} + d \bar{d}\big) ~ cos\phi_{V} - s \bar{s} ~ sin\phi_{V} \\
\omega & = & \frac{1}{\sqrt{2}} \big(u \bar{u} + d \bar{d}\big) ~ sin \phi_{V} - s \bar{s} ~ cos\phi_{V}
\end{eqnarray} 
with $\phi_{P(V)} = \theta_{ideal}-\theta_{P(V)} $,  $\theta_{ideal}= 35.3^{\circ}, ~\theta_{P} = -15.4^{\circ} ~\text{and}~ \theta_{V} = 40.3^{\circ}$ \cite{66}.

The decay constants used in the present work \cite{66,85,86,87}  are given below (in GeV), 
\begin{eqnarray}
f_{\pi} &=& 0.130, ~f_{K} = 0.155, ~ f_{D} = 0.211, 
f_{D_{s}} = 0.249,  ~ f_{B} = 0.190,~f_{B_{s}} = 0.227,  \nonumber  \\
f_{B^{*}} &=& 0.210,~  f_{B_{s}^{*}} = 0..251, ~ f_{D^{*}}  = 0.242, ~ f_{D_{s}^{*}} = 0.293. 
\end{eqnarray}
The radiative decay rate estimates the lifetime $\tau_{\bc}\sim 10^{-17}$ sec \cite{88,89}. We employ the factorization scheme to calculate the branching ratios for the two body nonleptonic weak decays of $ \bc $ ($\bsn$) meson to $ PP $ and $ PV $ decay modes using the modified BSW model. Obtained results are given in Tables \ref{t4}, \ref{t5},  \ref{t6}, \ref{t7}, \ref{t8} and \ref{t9}. We wish to remark that we have given the values of branching ratios only $\m(10^{-13})$ or larger. Furthermore, we have ignored the decays arising from the Class III transitions in $\bc(\bsn) \to PV$ decays as they involve contributions from nontrivial $\bc(\bsn) \to V$ form factors for being beyond the scope of present framework. We include the flavor dependent effects on average transverse quark momentum $ \omega $ inside the meson which will affect the form factors as well as the branching ratios. For a reasonable comparison, we also perform the QCD inspired calculation for $ \omega_{\Lambda}$ as an alternate point of view. We also compared our results with other recent works \cite{49,50,51,52}. We observe the following: 
 
\subsection{Branching ratios for \boldmath $ B_{q}^*$ decays ($q = c,~ s$)}
In the present analysis, branching ratios of $\bc$ and $\bsn$ decays including flavor dependent effects have been calculated. Once we compare numerical values of the flavor dependent $\bc(\bsn) \to P$ transition form factors with the values at fixed $\omega=0.40$ GeV, we find that the flavor dependent form factors are significantly enhanced. These form factors are further enhanced in QCD inspired case because of the large value of $ \omega_{\Lambda} $. One can expect a similar enhancement in the numerical branching ratios for the decays involving above said transitions. The results for branching ratios of $ \bc (\bsn) \to PP /PV $ decays are discussed as follows:
\subsection*{Bottom Conserving Modes}
\begin{itemize}
	\item[i.] The bottom conserving modes have enriched decay channels with the branching ratios $ \m(10^{-6})  \sim \m (10^{-9})$ due to the flavor dependent form factors and large values of CKM matrix elements involved as compared to the bottom changing decays. We have calculated the branching ratios for both $ \bcn \rightarrow PP $ and $ \bcn \rightarrow PV $ decays. As expected, the CKM-enhanced mode $ (\Delta b = 0, \Delta C = -1, \Delta S = -1)$ has the dominant branching ratios: $ \b(\bcn \rightarrow \bar{B}_s^0 \rho^-) = 9.25 \t 10^{-6} $; $ \b(\bcn \rightarrow B^- K^{*0}) = 1.35 \t 10^{-6} $; $ \b(\bcn \rightarrow \bar{B}_s^0 \pi^-) = 3.12 \t 10^{-7} $ and $ \b( \bcn \rightarrow B^- K^0) = 4.74 \t 10^{-8} $  for\textit{ PV} and \textit{PP} modes, respectively. The CKM elements $ V_{cs} $ and $ V_{du} $ plays a vital role in large branching ratios of these decays. The branching ratios for the Class I type decays: $\bcn \rightarrow \bar{B}_s^0 \pi^-$ and $ \bcn \rightarrow \bar{B}_s^0 \rho^- $ are further enhanced by an order of magnitude  owing to the color factor. It may be noted that the factorization results are considered to be very reliable for tree level color-favored modes. The ratio \[\frac{\b(\bcn \rightarrow \bar{B}_s^0 \rho^+)}{\b(\bcn \rightarrow \bar{B}_s^0 \pi^- )}= 29.7\]
	could be of immense interest for experimentalists. Since, the $B_c^- \rightarrow \bar{B}_s^0 \pi^-$ mode has already been seen \cite{66}, we have given the following ratios as another test of factorization and BSW model,   \[\frac{\b(\bcn \rightarrow \bar{B}_s^0 \rho^-)}{\b(B_c^- \rightarrow \bar{B}_s^0 \rho^- )}= 1.2\ten^{-4}~\text{and} ~\frac{\b(\bcn \rightarrow \bar{B}_s^0 \pi^-)}{\b(B_c^- \rightarrow \bar{B}_s^0 \pi^- )}= 5.9\ten^{-6};\] 
	here, we used $\b(B_c^+ \rightarrow \bar{B}_s^0 \rho^- )=4.84\%$ and $\b(B_c^- \rightarrow \bar{B}_s^0 \pi^- )=5.32\%$, respectively, from our previous work \cite{47,48}. 
	  
	On the other hand, the color-suppressed modes have always presented a challenge in understanding due to expected nonfactorizable contributions, therefore, we present \[
	\frac{\b(\bcn \rightarrow B^- K^{*0})}{\b(B_c^- \rightarrow B^- K^{*0})} = 1.68\ten^{-4} ~\text{and}~ \frac{\b(\bcn \rightarrow B^- K^{0})}{\b(B_c^+\rightarrow B^- K^0)} = 5.34\ten^{-6};\]
	using our previous results \cite{47,48}. These ratios between the dominant modes with pseudoscalar and vector meson final states would be helpful for the experimental searches.
	 
	\item[ii.]  Interestingly, the CKM-suppressed mode $ (\Delta b = 0, \Delta C = -1, \Delta S = 0) $ also have the branching ratios $ \m(10^{-7})  \sim \m (10^{-8})$ for the dominant decays \textit{i.e.},  $ \b(\bcn \rightarrow \bar{B}^0 \rho^-) = 4.24 \ten^{-7} $; $ \b(\bcn \rightarrow \bar{B}_s^0 K^{*-}) = 3.84 \ten^{-7} $; $ \b(\bcn \rightarrow \bar{B}_s^0 K^- )= 1.92 \t 10^{-8} $ and $ \b( \bcn \rightarrow \bar{B}^0 \pi^- )= 1.15 \t 10^{-8}  $. The color and kinematic factors overly compensate the CKM-suppression for these decay modes. It is essential to observe  that all the branching ratios are enhanced due to the flavor dependent effects. Further, the branching ratio of the most dominant decay, $ \bcn \rightarrow \bar{B}_s^0 \rho^-$, at fixed $ \omega = 0.40$ GeV yields $ 2.61 \t 10^{-6}$, which has increased by a factor of $ 3.5 $ for the flavor dependent case. The branching ratios for rest of the color-suppressed decays range in $ \m(10^{-9})  \sim \m (10^{-11})$. 
	The branching ratios for CKM-doubly-suppressed mode $ (\Delta b = 0, \Delta C = -1, \Delta S = 1)$ are highly suppressed \textit{i.e.} $\m(10^{-10})$. In addition to this, we have also given the following $SU(3) $ symmetry breaking relations \textit{w.r.t.} CKM-enhanced modes:
	\[
	\frac{ \b(\bcn \rightarrow  \bar{B}_s^0  K^{*-})}{ \b(\bcn \rightarrow \bar{B}_s^0 \rho^- )} = 0.041; ~\frac{ \b(\bcn \rightarrow  \bar{B}_s^0  K^{-})}{ \b(\bcn \rightarrow \bar{B}_s^0 \pi^- )} = 0.062; \]
	\[\frac{ \b(\bcn \rightarrow  B^-  K^{0})}{\b(\bcp \rightarrow B^- \pi^0 )} = 50.4;
	~\frac {\b(\bcp \rightarrow B^-   K^{*0} )} {\b(\bcp \rightarrow  B^-  \rho^{0})} = 38.6.
	\]
	It may be noted that we have ignored all the decays with branching ratios less than $ \m(10^{-13})$.
	\item[iii.]
	The major difference between QCD inspired calculation and flavor dependent calculation is due to the average transverse quark momenta $\omega $ which effects the form factors as well as branching ratios. As expected, in QCD inspired case all the values are enhanced by an order of magnitude due to larger $\omega_{\Lambda}$. The branching ratios of the most dominating decays are $ \b(\bcn \rightarrow \bar{B}_s^0 \rho^-) = 1.17 \t 10^{-5} $ and $ \b(\bcn \rightarrow B^- K^{*0}) = 2.31 \t 10^{-6} $ as shown in column 3 of Table \ref{t6}. Since, the distribution amplitudes for the QCD inspired $\omega_{\Lambda}$ are larger when compared to flavor dependent $\omega$ and the fixed $\omega=0.40$ GeV, the large overlap integrals results in enhanced form factors and therefore, branching ratios. Moreover, the branching ratios in case of QCD inspired results can be treated as an upper limit in the current analysis. 
	   
\end{itemize}
\subsection*{Bottom changing modes}
\begin{itemize}
	\item[i.] The branching ratios for bottom changing modes are suppressed. However, the decays involving the  $\bc \rightarrow \eta_c $ transition are enhanced because of the larger wave function overlap (see Fig \ref{fig2}) in flavor dependent form factors. In bottom changing $(\Delta b = 1)$ CKM-enhanced modes, $ ( \Delta C = 0, \Delta S = -1 )$ and $ (\Delta C = 1, \Delta S = 0)$, the branching ratios of the dominant decays are:  $ \b(\bcn \rightarrow \eta_c  \rho^-) = 7.72 \t 10^{-8} $; $ \b(\bcn \rightarrow \eta_c  D_s^-) = 1.58 \t 10^{-8} $ and $ \b(\bcn \rightarrow \eta_c \pi^{-}) = 7.69 \t 10^{-9}  $. Here, $ \bcn \rightarrow \eta_c  D_s^- $ is Class III type decay that get contributions from both color-favored and color-suppressed diagrams. However, the decays $\bcn \rightarrow \eta_c \pi^{-}$ and $ \bcn \rightarrow \eta_c  \rho^- $ receive contributions from color-favored diagrams only. To test the reliability of  framework, in the present case, we give the following ratios, 
	\[\frac{\b(\bcn \rightarrow \eta_{c}  \rho^-)}{\b(B_c^- \rightarrow \eta_{c}  \rho^-)} = 1.80\ten^{-5};\]
	\[\frac{\b(\bcn \rightarrow \eta_{c}  \pi^+)}{\b(B_c^- \rightarrow \eta_{c}  \pi^-)} = 5.00\ten^{-6};\]
	\[\frac{\b(\bcn \rightarrow \eta_{c} D_s^-)}{\b(B_c^- \rightarrow \eta_{c}  D_s^-)} = 2.74\ten^{-6}, \]
	using branching ratios from our previous work \cite{47,48}. It is worth mentioning that the results from previous work \cite{47,48}, in a similar framework, yields the following symmetry breaking ratios, 
	\[\frac{\b(\bcn \rightarrow J/\psi K^-)}{\b(\bcn \rightarrow J/\psi \pi^-)} = 0.076 ~~(Expt.: 0.079 \pm 0.007 \pm 0.003 );\]
	\[\frac{\b(\bcn \rightarrow J/\psi D_s^-)}{\b(\bcn \rightarrow J/\psi \pi^-)} = 2.2 ~~(Expt.: 3.1 \pm 0.5 );\] which are in good agreement with the experimental results \cite{66}. Similarly, we give following SU(3) and SU(4) symmetry breaking relations based on the present results,
	 \[\frac{\b(\bcn \rightarrow \eta_{c} K^{*-})}{\b(\bcn \rightarrow \eta_{c} \rho^-)} = 0.054;~\frac{\b(\bcn \rightarrow \eta_{c} K^-)}{\b(\bcn \rightarrow \eta_{c} \pi^-)} = 0.073;\frac{\b(\bcn \rightarrow \eta_{c} D_s^-)}{\b(\bcn \rightarrow \eta_{c} \pi^-)} = 2.05.\]
	
	\item[ii.] As expected, the QCD inspired branching ratios are enhanced by an order of magnitude: $ \b(\bcn \rightarrow \eta_c  \rho^-) = 1.61 \t 10^{-7} $, $ \b(\bcn \rightarrow \eta_c  D_s^-) = 1.90 \t 10^{-8} $ and  $ \b(\bcn \rightarrow \eta_c \pi^{-}) = 1.50 \t 10^{-8} $. 
\item[iii.] CKM-suppressed and -doubly-suppressed decay modes have the branching ratios $ \m(10^{-10}) $ to $ \m(10^{-13}) $, except for the decay, $ \bcn \rightarrow \eta_c  K^{*-}$, having branching ratio $4.28 \t 10^{-9} $. In general, the branching ratios of all the $\bc$ decays are enhanced by an order of magnitude for the QCD inspired case.
\item[iv.] Now, we shift our focus to rare weak decays of $\bsn$ meson, in bottom changing modes, which are considered to be important for the new physics searches beyond the Standard Model. In CKM-enhanced ($ \Delta C = 1, \Delta S = 0 $) and  $ (\Delta C = 0, \Delta S = -1)$ modes , the dominant decays have branching ratios: $ \b(\abs \rightarrow D_{s}^{+} D_{s}^{*-}) = 3.20 \t 10^{-7} $; $ \b(\abs \rightarrow D_s^+ \rho^-) = 1.07 \t 10^{-7} $;  $ \b(\abs \rightarrow D_{s}^{+} D_{s}^{-}) = 2.93 \t 10^{-8} $; $ \b(\abs \rightarrow D_s^+ \pi^-) = 1.08 \t 10^{-8} $; $ \b(\abs \rightarrow \eta  J/\psi) = 4.23 \t 10^{-9} $ and $ \b(\abs \rightarrow K^0 D^{*0}) = 2.20 \t 10^{-9} $. The Class I type color-favored decays $\abs \rightarrow D_{s}^{+} D_{s}^{*-}$,  $ \abs \rightarrow D_s^+ \rho^-$ and $ \abs \rightarrow D_s^+ \pi^-$ have larger branching ratios due to larger magnitude of $ \abs \rightarrow D_s^+ $ transition form factor. The branching ratios of these decays are further enhanced in QCD inspired case. 
\item[v.]  CKM--suppressed and -doubly-suppressed decay modes for $\bsn$ meson have the branching ratios $ \m(10^{-10}) $ to $ \m(10^{-12})$ except $ \b(\abs \rightarrow D_{s}^{+} D^{*-}) = 1.10 \t 10^{-8} $ and $ \b(\abs \rightarrow D_s^{+} K^{*-}) = 5.81 \t 10^{-9} $. Once again, the larger $ \abs \rightarrow D_s^+ $ transition form factors along with color-favored diagram result in larger branching ratios for these modes. In order to see the effects of flavor dependence in $ \bsn $ decays, we compute results for $ \abs \rightarrow D_s^+ D_s^{*-} $  and $ \abs \rightarrow D_{s}^{+} D^{*-}$ decays at $ \omega = 0.40 $ GeV, which in turn yield the  branching ratios $ 8.66 \t 10^{-8} $ and  $ 8.57 \t 10^{-9} $, respectively. We observed that the flavor dependent branching ratios are enhanced by a factor of $ 3.6$ and $1.3$, respectively, for the above said decays. 
\item[vi.] It is reasonable to test the factorization hypothesis for color-suppressed decays  in $ \bsn $ decay channels. Here, we give the following ratios, 
\[
	\frac{ \b(\abs \rightarrow D_{s}^{+} D_s^{*-} )}{ \b(\abs \rightarrow  D_{s}^{+} D_s^{-})} = 10.92;~\;
	\frac{\b(\abs \rightarrow D_{s}^{+}\rho^-)} {\b(\abs \rightarrow  D_{s}^{+}\pi^- )} = 9.91.
\]
 which would be helpful for the future experimental observations. From the above ratios it is also clear that the branching ratios of $ \abs \rightarrow PV $ decays are considerably higher than that of $ \abs \rightarrow PP $ decays. Our analysis for SU(3) flavor symmetry in case of $\bsn$ decays yield 
\[
	\frac{\b(\abs \rightarrow D_{s}^{+} K^{*-} )}{\b(\abs \rightarrow D_{s}^{+} \rho^{-} )} = 0.0528.
\]
 Here again, the numerical results in QCD inspired calculation are higher than the flavor dependent results due to the larger values of $\omega_{\Lambda}$. 
\end{itemize}
We have also compared our results with the recent predictions from other works, which are listed in the Tables \ref{t5}, \ref{t7}, and \ref{t8}. It is interesting to note that our results for QCD inspired case are of the same order as compared to J. Sun {\it et al.} \cite{49} for their $ \omega = \alpha_{s} m $ case. We wish to point out that the quark masses used by \cite{49} are substantially different from our inputs, especially $ m_{c} $ and $ m_{b} $. Also, the Wilson's coefficients used in both the works are different. Moreover, $ \omega _{\Lambda} $ in our case for $ \bc $ and $ B $ systems are $ 1.33 $ and $ 1.12 $, respectively. However, their results are on larger side when compared with our flavor dependent work, owing to the large values of form factors corresponding to their inputs. In bottom changing decays, our results are comparable with the perturbative QCD framework \cite{51} \textit{i.e.}  $ \b( \bcn \rightarrow \eta_c \pi^-) =  2.22 \ten^{-8} $ and  $ \b( \bcn \rightarrow  \eta _c K^-) = 1.67 \ten^{-9} $. We also compare our results with the branching ratios $ \b(\bcn \rightarrow \eta_c \rho^-) = 3.02 \t 10^{-8} $, $ \b(\bcn \rightarrow \eta_c K^{*-}) = 1.7 \t 10^{-9} $ calculated in QCDF \cite{50} and $ \b(\bcn \rightarrow \eta_c \rho^-) = 2.45 \t 10^{-8} $, $ \b(\bcn \rightarrow \eta_c K^{*-}) = 1.4 \t 10^{-9} $ in LFQM \cite{50}, are of the same order. Similar studies has also been done for the rare weak decays of $ \bsn $ in the factorization approach \cite{52} and the branching ratios in the present work are, in general, larger as compared to their results. It is worth pointing out that many of the results from decay channels under study, in different models, are well within the reach of current experiments and future experimental searches. 
\section{Summary and Conclusions}
In the present work, we have calculated the branching ratios of rare weak decays of $ \bc $ and $ \bsn $ mesons using the modified BSW model framework. The detailed analysis of  $ \bc \rightarrow PP,PV $ decays for both bottom conserving and bottom changing modes and $ \bsn\rightarrow PP,PV $ decays for bottom changing mode has been presented. We have included the flavor dependent effects on average transverse quark momenta inside the meson that affects the form factor and consequently, the branching ratios of these decays. In addition to the flavor dependence analysis, we have also performed a study based on the QCD inspired approximation of transverse quark momentum, $\omega_{\Lambda}$. We summarize our findings as follows:
\begin{itemize}
	\item[i.] It is observed that the flavor dependent numerical values of $\omega$ have significantly enhanced the form factors and the branching ratios of $\bc(\bsn)$ decays. The branching ratios of these decay channels are further enhanced for the QCD inspired form factors due to larger values of $\omega_{\Lambda}$.
	
	\item[ii.] As expected, $ \bc \rightarrow PV $ decays have larger branching ratios as compared to $\bc \to PP$ decays, therefore, they are the most probable candidates for experimental detection. The branching ratios of most dominant $\bc \to PV$ decays, namely, $ \bcp \rightarrow B_{s}^{0} \rho^{+}$ /$ B^{+} \bar{K^{*0}}$ are of $ \m (10^{-6}) $ in the flavor dependent case. However, these branching ratios are further enhanced by an order to $ \m (10^{-5})$ for the QCD inspired case. 
	 
	\item[iii.] A number of decay channels have the flavor dependent branching ratios $ \m(10^{-7})$ to $ \m(10^{-8})$, which are further enhanced due to QCD inspired form factors. Further, the number of possible weak decay channels of $\bc$ meson having branching ratios $\gtrsim \m(10^{-8})$ are large enough to grab the attention of experimentalists.  
	
	\item[iv.] In case of $\bsn$ decays, branching ratios of $ B_{s}^{*} \rightarrow PV $ are relatively larger than the $ B_{s}^{*} \rightarrow PP $ decays with the dominating channels $ \b(\abs \rightarrow D_s^+ \rho^-) = 1.10 \t 10^{-7} $; $ \b(\abs \rightarrow D_{s}^{+} D_{s}^{*-}) = 3.27 \t 10^{-7} $. Here also, the QCD inspired branching ratios are larger as compared to the flavor dependent case. The $\bsn$ decays are expected to be important for the new physics scenario, however, they have not yet been studied much in detail. 
\end{itemize}

The observation of the $\bc$ meson is stalled, mainly, due to the difficulties in its production and partly, due to the high electromagnetic background of LHC. However, it is only a matter of time that this important particle will be observed by LHCb in near future. We hope that our results on rare weak decay channels could play crucial role in observation and studies of heavy flavor vector mesons.

\newpage

\newpage


\newpage
\begin{table}[h]
	\caption {The $ \bc \to P $ transition form factors.}
	\label{t1}
	\begin{tabular}{ |c|c|c|c|c|c|}
		\hline
		Decay & This work & $V$ & $A_{0}$ & $A_{1}$ & $ A_{2}$ \\
		\hline
		$ \bc \rightarrow B_{s} $  & at $\omega=0.40$ GeV  & 2.890 & 0.368 & 0.433 & 0.428\\
		& Flavor dependent & $ 5.560$ & $0.582$ & $0.833$ & $ 2.466$ \\
		& QCD inspired & 6.516 & 0.920 & 0.976 & -0.240 \\
		
		\hline
		$ \bc \rightarrow B $ 	& at $\omega=0.40$ GeV & 2.624  & 0.305 & 0.345 & 0.133\\
		& Flavor dependent  & $ 5.086 $ & $0.431$ & $0.668$ & $ 2.189$ \\
		& QCD inspired & 6.927  & 0.903 & 0.910 & -0.833\\
	
		\hline	
		$ \bc \rightarrow \eta_{c} $ & at $\omega=0.40$ GeV & 0.205 & 0.147 & 0.148 & -0.147\\
		 & Flavor dependent & $ 0.793 $ & $ 0.526 $ & $ 0.571$ & -$0.400$ \\
		& QCD inspired & 1.214 & 0.730 & 0.874 & -0.331\\
		
		\hline 
		$ \bc \rightarrow D_{s} $  & at $\omega=0.40$ GeV & 0.025 & 0.017 & 0.016 & -0.019\\
		& Flavor dependent & $ 0.180 $ & $ 0.104 $ & $ 0.115$ & -$ 0.082 $ \\
		& QCD inspired & 0.672 &  0.342 & 0.429& -0.176\\
		
		\hline
		$ \bc \rightarrow D $  & at $\omega=0.40$ GeV & 0.020  & 0.013 & 0.012 & -0.016\\
		& Flavor dependent & $ 0.084 $ & $ 0.0451$ & $ 0.052 $ & -$ 0.033$ \\
		& QCD inspired & 0.598  & 0.292 & 0.368& -0.153\\
		
		\hline
		$ \bc \rightarrow \pi $  & at $\omega=0.40$ GeV & 0.335 & 0.473 & 0.643 & -0.296\\
		& Flavor dependent & $ 0.319 $ & $ 0.416 $ & $ 0.613 $ & -$ 0.211 $ \\
		& QCD inspired & 0.426 & 0.490 & 0.817 & -0.149\\

		\hline
		$ \bc \rightarrow \eta $  & at $\omega=0.40$ GeV & 0.352 & 0.463 & 0.592 & -0.309\\
		& Flavor dependent & $ 0.254 $ & $ 0.302 $ & $ 0.427 $ & -$ 0.154 $ \\
		& QCD inspired & 0.463 & 0.496 & 0.780 & -0.157\\
		
		\hline
		$ \bc \rightarrow \eta^{\prime} $  & at $\omega=0.40$ GeV & 0.359 & 0.436 & 0.529 & -0.310\\
		& Flavor dependent & $ 0.131 $ & $ 0.131 $ & $ 0.193 $ & -$0.047 $ \\
		& QCD inspired & 0.605 & 0.598 & 0.892 & -0.199\\
		
		\hline
		
	\end{tabular}
\end{table}

\begin{table}[h]
	\caption {The $ \bsn \to P $ transition form factors.}
	\label{t2}
	\begin{tabular}{ |c|c|c|c|c|c|}
		\hline 
		Decay & This work & $V$ & $A_{0}$ & $A_{1}$ & $ A_{2}$ \\
		\hline
		$ \bsn \rightarrow D_{s} $  & at $\omega=0.40$ GeV & 0.676 & 0.605 & 0.632 & -0.546\\
		& Flavor dependent & $ 0.774 $ & $ 0.671 $ & $ 0.724 $ & -$ 0.559 $ \\
		& QCD inspired & 1.022 & 0.770 & 0.955 & -0.370\\
		
		\hline
		$ \bsn \rightarrow K $  & at $\omega=0.40$ GeV & 0.295 & 0.268 & 0.277& -0.254\\
		& Flavor dependent & $ 0.273 $ & $0.237$ & $0.256$ & -$ 0.213$ \\
		
		& QCD inspired & 0.657  &  0.477 & 0.618 & -0.307\\

		\hline
		$ \bsn \rightarrow \pi $  & at $\omega=0.40$ GeV & 0.334 & 0.339 & 0.387 & -0.289\\
		& Flavor dependent & $ 0.354$ & $0.347$ & $0.410$ & -$ 0.281$ \\
		
		& QCD inspired & 0.564 & 0.462 & 0.652 & -0.262 \\
		
		\hline
		$ \bsn \rightarrow \eta $  & at $\omega=0.40$ GeV & 0.335 & 0.313 & 0.333 & -0.288\\
		& Flavor dependent & $ 0.250 $ & $ 0.223 $ & $ 0.248 $ & -$ 0.191 $ \\
		& QCD inspired & 0.613  & 0.460 & 0.608 & -0.279 \\
		
		\hline
		$ \bsn \rightarrow \eta^{\prime} $  & at $\omega=0.40$ GeV & 0.306 & 0.259 & 0.260 & -0.259\\
		& Flavor dependent & $ 0.043 $ & $ 0.032 $ & $ 0.036 $ & -$ 0.026 $ \\
		& QCD inspired & 0.785  & 0.537 & 0.667 & -0.350\\
		
		\hline
		
	\end{tabular}
\end{table}
\begin{table}[h]
	\caption {$ |\Psi(0)|^{2} $ and $ \omega $ values for vector and pseudoscalar mesons }
	\label{t3}
	\begin{tabular}{ |c|c|c|c|}
		\hline
		Meson & $ |\Psi(0)|^{2} $ & Flavor dependent $ \omega $ & QCD inspired $ \omega_{\Lambda} $ \\ \
		& (GeV$^{3}$) &  (GeV) &  (GeV) \\
		\hline  
		$ K^{*} (K) $ & 0.009 & 0.30 & 0.54 \\
		\hline
		$ \pi (\rho) $ & 0.009 & 0.30 & 0.46 \\
		\hline 
		$ D^{*} (D) $ & 0.021 & 0.41 & 0.62 \\
		\hline 
		$ D_{s}^{*} (D_{s}) $ & 0.034 & 0.47 & 0.65\\
		\hline 
		$ J/\psi (\eta_{c}) $ & 0.094 & 0.67 & 0.96 \\
		\hline 
		$ B^{*} (B) $ & 0.030 & 0.45 & 1.12 \\
		\hline 
		$ B_{s}^{*}(B_{s})$ & 0.049 & 0.53 & 1.14 \\
		\hline 
		$ B_{c}^{*} (B_{c})$ & 0.196 & 0.85 & 1.33 \\
		\hline 
	\end{tabular}
\end{table}

\begin{center}
	\begin{table}[h]
		\caption {Branching ratios of $B_c^{*+} \rightarrow PP $ decays for CKM-enhanced and -suppressed modes.}
		\label{t4}
		\begin{tabular}{|c|c|c|c|}
			\hline
			\tr & Flavor dependent & QCD inspired & \cite{49} \\
			\hline
			\multicolumn{4}{|c|}{\ce($ \Delta b = 0, \Delta C = -1, \Delta S = -1$)}\\ 
			\hline
			$  \bcn \rightarrow \bar{B}_s^0 \pi^- $ & $3.12 \ten^{-7}$  &  $ 7.78 \ten^{-7} $  & $ 9.82 \ten^{-7} $ \\
			\hline
			$ \bcn \rightarrow B^- K^0 $ &$4.74 \ten^{-8}$  &  $ 2.08 \ten^{-7} $  & $ 1.06 \ten^{-7} $ \\
			\hline
			
			\multicolumn{4}{|c|}{\cs($ \Delta b = 0, \Delta C = -1, \Delta S = 0 $)}\\
			\hline
			$ \bcn \rightarrow \bar{B}^0 \pi^- $ & $ 1.15 \ten^{-8} $ &  $ 5.04  \ten^{-8} $ & $ 6.53 \ten^{-8} $ \\
			\hline
			$ \bcn \rightarrow \bar{B}_s^0 K^- $ & $ 1.92 \ten^{-8} $ &  $ 4.78 \ten^{-8} $ & $ 4.87 \ten^{-8} $ \\
			\hline
			$ \bcn \rightarrow B^- \eta $ & $2.45 \ten^{-9} $ &  $ 1.07 \ten^{-8} $ & $ 7.22 \ten^{-9} $\\
			\hline
			$  \bcn \rightarrow B^- \pi^0 $ & $9.41 \ten^{-10}  $ &  $ 4.13 \ten^{-9} $ & $ 2.82 \ten^{-9} $ \\
			\hline
			
			$ \bcn \rightarrow B^- \eta^{\prime} $ & $4.44 \ten^{-11}$ &  $ 1.95 \ten^{-10} $ & $ 2.02 \ten^{-10} $  \\
			\hline
			\multicolumn{4}{|c|}{\cds ($ \Delta b = 0, \Delta C = -1, \Delta S = 1 $)}\\
			\hline
			$ \bcn \rightarrow \bar{B}^0 K^{-} $ & $7.85 \ten^{-10}$ &  $ 3.45 \ten^{-9} $ & $ 3.50 \ten^{-9} $ \\
			\hline
			$\bcn \rightarrow B^- \bar{K}^0 $ & $1.29 \ten^{-10} $ &  $ 5.67 \ten^{-10} $ & $ 3.00 \ten^{-10} $ \\
			\hline

		\end{tabular}
	\end{table}
	
\begin{table}[h]
	
	\captionof{table} {Branching ratios of $B_c^{*-} \rightarrow PP $ decays for CKM-enhanced and -suppressed modes.}
	\label{t5}
	\begin{tabular}{|c|c|c|}
		\hline
		\tr & Flavor dependent & QCD inspired  \\
		\hline
		\multicolumn{3}{|c|}{\ce($ \Delta b = 1, \Delta C = 0, \Delta S = -1 $) }\\
		\hline
		$ \bcn \rightarrow \eta_c  D_s^- $ & $1.58 \ten^{-8}$ &  $ 1.96  \ten^{-8} $ \\
		\hline
		
		\multicolumn{3}{|c|}{\ce($ \Delta b = 1, \Delta C = 1, \Delta S = 0$)}\\
		\hline 
		$  \bcn \rightarrow \eta_c \pi^- $ & $ 7.70\ten^{-9} $ &  $ 1.48 \ten^{-8} $ \\
		\hline
		$  \bcn \rightarrow D^- D^0 $  & $1.55 \ten^{-11} $&  $ 6.47  \ten^{-10} $  \\
		\hline 
		
		\multicolumn{3}{|c|}{\cs($ \Delta b = 1, \Delta C = 0, \Delta S = 0 $ )}\\
		\hline 
		$  \bcn \rightarrow \eta_c  D^- $ & $ 7.09 \ten^{-10} $ &  $ 7.36 \ten^{-10} $  \\
		\hline 
		$  \bcn \rightarrow \bar{D^0} \pi^-$ & $ 7.25 \ten^{-13} $ &  $ 3.05  \ten^{-11} $  \\
		\hline 
		\multicolumn{3}{|c|}{\cs($ \Delta b = 1, \Delta C = 1, \Delta S = -1$)}\\
		\hline 
		$  \bcn \rightarrow  \eta _c K^- $ & $5.66 \ten^{-10}$ &  $ 1.10 \ten^{-9} $   \\
		\hline 
		$  \bcn \rightarrow  D_s^- D^0 $ & $4.00 \ten^{-12}$ &  $ 4.34 \ten^{-11} $  \\
		\hline 
		\multicolumn{3}{|c|}{\cs($ \Delta b = 1, \Delta C = -1, \Delta S = -1$)}\\
		\hline
		$ \bcn \rightarrow D_s^{-} \bar{D^0} $ & $1.01 \ten^{-12} $  &  $ 8.11 \ten^{-11} $ \\
		\hline
		\multicolumn{3}{|c|}{\cds($ \Delta b = 1, \Delta C = -1, \Delta S = 0$)}  \\
		\hline
		$ \bcn \rightarrow \bar{D^0} D^- $ & $6.95 \ten^{-14}$ &  $ 2.93 \ten^{-12} $ \\
		\hline
	\end{tabular}
\end{table}

		\begin{table}[h]
		\caption {Branching ratios of $B_c^{*+} \rightarrow PV $ decays for CKM-enhanced and -suppressed modes.}
		\label{t6}
		\begin{tabular}{ |c|c|c|c|}
			\hline
			\tr & Flavor dependent & QCD inspired & \cite{49}\\
			\hline
			\multicolumn{4}{|c|}{\ce($ \Delta b = 0, \Delta C = -1, \Delta S =-1$)}\\ 
			\hline
			$ \bcn \rightarrow \bar{B}_s^0 \rho^- $ & $ 9.25 \ten^{-6}$&  $ 1.37  \ten^{-5} $ & $1.67 \ten^{-6}$  \\
			\hline
			$ \bcn \rightarrow B^- K^{*0} $ & $1.35 \ten^{-6}$ &  $ 2.70  \ten^{-6} $ & $1.88 \ten^{-7}$\\
			\hline
			
			\multicolumn{4}{|c|}{\cs($ \Delta b = 0, \Delta C = -1, \Delta S = 0 $)} \\ 
			\hline
			$ \bcn \rightarrow \bar{B}_s^0 K^{*-}  $  & $ 3.84 \ten^{-7}$  &  $5.42  \ten^{-7} $ & $6.01 \ten^{-8}$\\
			\hline
			$ \bcn \rightarrow  \bar{B}^0 \rho^- $  & $4.24 \ten^{-7}$ &  $ 9.06  \ten^{-7} $ & $1.15 \ten^{-7}$ \\
			\hline
			$ \bcn \rightarrow B^- \phi $  & $4.08 \ten^{-8} $ &  $ 7.69  \ten^{-8} $ & $4.04 \ten^{-9}$ \\
			\hline
			$ \bcn \rightarrow B^- \rho^0 $ & $ 3.47 \ten^{-8}$&  $ 7.43 \ten^{-8} $ & $6.08 \ten^{-9}$ \\
			\hline 
			$ \bcn \rightarrow B^- \omega $ & $3.12 \ten^{-8} $ &  $ 6.64  \ten^{-8} $ &  $4.51 \ten^{-9}$\\
			\hline
			\multicolumn{4}{|c|}{\cds($ \Delta b = 0, \Delta C = -1, \Delta S = 1 $)} \\ 
			\hline
			$ \bcn \rightarrow \bar{B}^0 K^{*-} $ & $2.23 \ten^{-8}$&  $ 4.48  \ten^{-8} $ & $4.84 \ten^{-9}$ \\
			\hline 
			$ \bcn \rightarrow B^- \bar{K}^{*0} $ & $3.67 \ten^{-9}$ &  $ 7.35 \ten^{-9} $ & $5.34 \ten^{-10}$ \\
			\hline

		\end{tabular}
	\end{table}
	
	\begin{table}[h]
		
		\caption {Branching ratio of $B_c^{*-} \rightarrow PV $ decays for CKM-enhanced and -suppressed modes.}
		\label{t7}
		\begin{tabular}{ |c|c|c|}
			\hline
			\tr & Flavor dependent & QCD inspired \\
			\hline
			\multicolumn{3}{|c|}{\ce($ \Delta b = 1, \Delta C = 1, \Delta S = 0$)} \\ 
			\hline
			$ \bcn \rightarrow \eta_c \rho^- $ & $ 7.72 \t 10^{-8} $ & $ 1.54 \ten^{-7} $ \\
			\hline
			$ \bcn \rightarrow D^- D^{*0} $ & $ 1.20 \t 10^{-10} $ & $ 5.60 \ten^{-9} $  \\
			\hline 
			\multicolumn{3}{|c|}{\ce ($ \Delta b = 1, \Delta C = 0, \Delta S = -1$)} \\
			\hline
			$ \bcn \rightarrow \bar{D}^0 K^{*-} $ & $4.00 \ten^{-13}$ &  $ 1.74  \ten^{-11} $ \\
			\hline
			\multicolumn{3}{|c|}{ \cs($\Delta b = 1, \Delta C = 1, \Delta S = -1$)} \\ 
			\hline
			$ \bcn \rightarrow \eta_c K^{*-} $ & $ 4.18 \t 10^{-9} $ & $ 8.44 \ten^{-9} $  \\
			\hline
			$ \bcn \rightarrow D_s^- D^{*0} $ & $ 3.00 \t 10^{-11} $ & $ 3.76 \ten^{-10}$   \\
			\hline
			\multicolumn{3}{|c|}{\cs($ \Delta b = 1, \Delta C = 0, \Delta S = 0$)} \\
			\hline
        	$ \bcn \rightarrow \bar{D}^0 \rho^- $ & $ 7.34 \ten^{-12}$ &  $ 3.18 \ten^{-10} $ \\
			\hline
			$ \bcn \rightarrow D^- \rho^0 $ & $2.00 \ten^{-13}$&  $ 8.57 \ten^{-12} $  \\
			\hline
			$ \bcn \rightarrow D^- \phi $ & $2.19 \ten^{-13}$&  $ 9.57 \ten^{-12} $  \\
			\hline
			
		\end{tabular}
	\end{table}

	\begin{table}[h]
		\captionof{table} {Branching ratios of $ \bs \rightarrow PP $ decays for CKM-enhanced and -suppressed modes.}
		\label{t8}
		\begin{tabular}{ |p{3cm}|p{3cm}|p{3cm}|p{3cm}|}
			\hline
			\tr & Flavor dependent & QCD inspired & \cite{52} \\
			\hline
			\multicolumn{4}{|c|}{\ce($ \Delta b = 1, \Delta C = 0, \Delta S = -1$)} \\
			\hline
			$\abs \rightarrow D_{s}^{-} D_{s}^{+}$ & $ 2.94 \t 10^{-8} $ & $ 3.85 \ten^{-8} $ & $ 2.4 \t 10^{-8} $ \\
			\hline 
			$\abs \rightarrow \eta_c \eta $ & $ 5.07 \t 10^{-10} $ & $ 2.17 \ten^{-9} $ & -\\
			\hline
			$\abs \rightarrow \eta_c \eta^{\prime} $ & $ 1.31 \t 10^{-11} $ & $ 3.65 \ten^{-9} $ & - \\
			\hline
			$\abs \rightarrow K^{+} K^{-}$ & $ 1.21 \t 10^{-12} $ & $ 4.88 \ten^{-12} $ & - \\
			
			\hline
			\multicolumn{4}{|c|}{\ce($ \Delta b = 1, \Delta C = 1, \Delta S = 0$)} \\
			\hline
			$\abs \rightarrow D_s^+ \pi^- $  & $ 1.09 \t 10^{-8} $ & $ 1.43 \ten^{-8} $ & $4.6 \ten^{-9}$\\
			\hline
			$\abs \rightarrow D^0 K^0 $ & $ 3.33 \t 10^{-10}$ & $ 1.35 \ten^{-10} $ & $ 1.5 \ten^{-10} $\\
			\hline
	
			\multicolumn{4}{|c|}{\cs($ \Delta b = 1, \Delta C = 0, \Delta S = 0$)} \\
			\hline
			$\abs \rightarrow D^- D_s^+ $ & $ 1.15 \t 10^{-9} $ & $ 1.51\ten^{-9} $ & $ 8.6 \t 10^{-10} $ \\
			\hline
			$\abs \rightarrow \eta_c K^0 $ & $ 8.05 \t 10^{-11} $  & $ 3.27 \ten^{-10} $ & -\\
			\hline
			$\abs \rightarrow \pi^0 K^0 $ & $ 4.31 \t 10^{-13} $ & $ 1.75 \ten^{-12} $ & -  \\
			\hline
			$\abs \rightarrow \pi^- K^+ $ & $ 1.61 \t 10^{-11} $ & $ 6.51 \ten^{-11} $ & -  \\
			\hline
			
			\multicolumn{4}{|c|}{\cs($ \Delta b = 1, \Delta C = 1, \Delta S = -1$)} \\
			\hline
			$\abs \rightarrow D_s^+ K^- $ & $ 8.00 \t 10^{-10}$ & $ 1.05 \ten^{-9} $ & $ 8.7 \t 10^{-10} $\\
			\hline
			$\abs \rightarrow D^0 \eta $ & $ 6.00 \t 10^{-12}$  & $ 2.56 \ten^{-11} $ & -\\
			\hline
			$\abs \rightarrow D^0 \eta ^{\prime} $ & $ 1.69 \t 10^{-13}$ & $ 4.73 \ten^{-11} $ & - \\
			\hline
	
			\multicolumn{4}{|c|}{\cds($ \Delta b = 1, \Delta C = -1, \Delta S = -1$)} \\
			\hline
			$\abs \rightarrow D_s^- K^+ $ & $ 6.88 \t 10^{-11} $  & $ 2.79 \ten^{-10} $ & $ 5.9 \ten^{-11} $ \\
			\hline
			$\abs \rightarrow \bar{D^0} \eta $ &  $ 9.09 \t 10^{-13} $  & $ 3.89 \ten^{-12} $ & - \\ 
			
			\hline 
			\multicolumn{4}{|c|}{\cds($ \Delta b = 1, \Delta C = -1, \Delta S = 0$)} \\
			\hline
			$\abs \rightarrow D^- K^+ $ & $ 2.56 \t 10^{-12} $  & $ 1.04 \ten^{-11} $ & $2.1\ten^{-12} $ \\
			\hline
			$\abs \rightarrow \bar{D^0} K^0 $ & $ 1.37 \t 10^{-13} $  & $ 5.57 \ten^{-13} $ & $ 6.0 \t 10^{-14} $\\
			\hline
		\end{tabular}
	\end{table}
	
	\begin{table}[h]
		
		\captionof{table} {Branching ratios of $ \bs \rightarrow PV $ decays for CKM-enhanced and -suppressed modes.}
		\label{t9}
		\begin{tabular}{ |p{3cm}|p{3cm}|p{3cm}|}
			\hline
			\tr & Flavor dependent & QCD inspired \\
			\hline
			
			\multicolumn{3}{|c|}{\ce ($ \Delta b = 1, \Delta C = 0, \Delta S = -1$)} \\ 
			\hline
		    $ \abs \rightarrow D_{s}^{+} D_{s}^{*-} $ & $ 3.20 \t 10^{-7} $  & $ 5.14 \ten^{-7}$ \\
			\hline
			$ \abs \rightarrow \eta J/ \Psi $ & $ 4.24 \t 10^{-9} $  & $ 2.37 \ten^{-8}$  \\
			\hline
		    $ \abs \rightarrow \eta^{\prime} J/ \Psi $ & $ 1.34 \t 10^{-10}$ & $ 4.34 \ten^{-8} $ \\
			\hline
			$ \abs \rightarrow K^{+} K^{*-} $ & $ 8.39 \t 10^{-12} $   & $ 3.62 \ten^{-11}$ \\
			\hline
			 
		\multicolumn{3}{|c|}{\ce($ \Delta b = 1, \Delta C = 1, \Delta S = 0 $)} \\ 
		\hline
		$ \abs \rightarrow D_s^+ \rho^- $ & $ 1.08 \t 10^{-7} $ & $ 1.49 \ten^{-7} $ \\
		\hline
		$ \abs \rightarrow K^{0} D^{*0}$ & $ 2.21 \t 10^{-9} $ & $ 1.09 \ten^{-8} $ \\
		\hline
		
			\multicolumn{3}{|c|}{\cs($ \Delta b = 1, \Delta C = 0, \Delta S = 0$)} \\ 
			\hline
			$ \abs \rightarrow D_{s}^{+} D^{*-}$ & $ 1.10 \t 10^{-8} $ & $ 1.75 \ten^{-8} $ \\
			\hline
			$ \abs \rightarrow K^+ \rho^-  $ & $ 1.56 \t 10^{-10} $ & $ 6.64 \ten^{-10} $ \\
			\hline
			$ \abs \rightarrow K^0 \rho^0 $ & $ 4.19 \t 10^{-12}$ & $ 1.79 \ten^{-11} $ \\
			\hline
			$ \abs \rightarrow K^0 \phi $ & $ 4.48 \t 10^{-12}$ & $ 1.97 \ten^{-11} $ \\
			\hline
			$ \abs \rightarrow K^0 \omega $ & $ 2.63 \t 10^{-13}$ & $ 1.12 \ten^{-12} $ \\
			\hline
			$ \abs \rightarrow K^0 J/\Psi $ & $ 6.40 \t 10^{-10} $ & $ 3.45 \ten^{-9} $ \\
			\hline
			\multicolumn{3}{|c|}{\cs($ \Delta b = 1, \Delta C = 1, \Delta S = -1$)} \\ 
			\hline
			$ \abs \rightarrow D_s^{+} K^{*-}$ & $ 5.82 \t 10^{-9} $ & $ 8.17 \ten^{-9} $ \\
			\hline
			$ \abs \rightarrow \eta D^{*0}  $ & $ 4.06 \t 10^{-11} $ & $ 2.09 \ten^{-9} $  \\
			\hline
			$ \abs \rightarrow \eta^{\prime} D^{*0}  $ & $ 1.29 \t 10^{-12} $ & $ 4.02 \ten^{-10} $  \\
			\hline
			
		    \multicolumn{3}{|c|}{\cds($ \Delta b = 1, \Delta C = -1, \Delta S = -1$)} \\ 
			\hline
			$ \abs \rightarrow K^{+} D_{s}^{*-} $ & $ 5.05 \t 10^{-10} $ & $ 2.51 \ten^{-9} $\\
			\hline
			$ \abs \rightarrow \eta \bar{D^{0*}} $ & $ 6.17 \t 10^{-12} $ & $ 3.18 \ten^{-11} $  \\
			\hline
			$ \abs \rightarrow \eta^{\prime} \bar{D^{0*}} $ & $ 1.97 \t 10^{-13} $ & $ 6.10 \ten^{-11} $  \\
			\hline

			\multicolumn{3}{|c|}{\cds($ \Delta b = 1, \Delta C = -1, \Delta S = 0$)} \\ 
			\hline
			$ \abs \rightarrow K^{+} D^{*-} $ & $ 1.70 \t 10^{-11} $ & $ 8.36 \ten^{-11} $  \\
			\hline
				$ \abs \rightarrow K^{0} \bar{D}^{*0} $ & $ 9.11 \t 10^{-13} $ & $ 4.50 \ten^{-12} $  \\
			\hline
						
   \end{tabular}
	\end{table}

\end{center}


\begin{thebibliography}{99}
	\section*{REFERENCES}
	\bibitem{1}
	F.~Abe {\it et al.} [CDF Collaboration],
	Phys.\ Rev.\ D {\bf 58} (1998) 112004
	doi:10.1103/PhysRevD.58.112004
	[hep-ex/9804014].
	\bibitem{2} 
	K.~Ackerstaff {\it et al.} [OPAL Collaboration],
	Phys.\ Lett.\ B {\bf 420}, 157 (1998)
	doi:10.1016/S0370-2693(97)01569-4
	[hep-ex/9801026].
	\bibitem{3}
	R.~Aaij {\it et al.} [LHCb Collaboration],
	Phys.\ Rev.\ Lett.\  {\bf 109}, 232001 (2012)
	doi:10.1103/PhysRevLett.109.232001
	[arXiv:1209.5634 [hep-ex]].  
	\bibitem{4}
	R.~Aaij {\it et al.} [LHCb Collaboration],
	Phys.\ Rev.\ Lett.\  {\bf 108}, 251802 (2012)
	doi:10.1103/PhysRevLett.108.251802
	[arXiv:1204.0079 [hep-ex]].  
	\bibitem{5}
	R.~Aaij {\it et al.} [LHCb Collaboration],
	Phys.\ Rev.\ Lett.\  {\bf 111}, no. 18, 181801 (2013)
	doi:10.1103/PhysRevLett.111.181801
	[arXiv:1308.4544 [hep-ex]].  
	\bibitem{6}
	R.~Aaij {\it et al.} [LHCb Collaboration],
	Phys.\ Rev.\ D {\bf 87}, 071103 (2013)
	doi:10.1103/PhysRevD.87.071103
	[arXiv:1303.1737 [hep-ex]].  
	\bibitem{7}
	R.~Aaij {\it et al.} [LHCb Collaboration],
	JHEP {\bf 1309}, 075 (2013)
	doi:10.1007/JHEP09(2013)075
	[arXiv:1306.6723 [hep-ex]].  
	\bibitem{8}
	R.~Aaij {\it et al.} [LHCb Collaboration],
	JHEP {\bf 1311}, 094 (2013)
	doi:10.1007/JHEP11(2013)094
	[arXiv:1309.0587 [hep-ex]].  
	\bibitem{9}
	R.~Aaij {\it et al.} [LHCb Collaboration],
	Eur.\ Phys.\ J.\ C {\bf 74}, no. 5, 2839 (2014)
	doi:10.1140/epjc/s10052-014-2839-x
	[arXiv:1401.6932 [hep-ex]]. 
	\bibitem{10}
	R.~Aaij {\
		it et al.} [LHCb Collaboration],
	Phys.\ Rev.\ Lett.\  {\bf 113}, no. 15, 152003 (2014)
	doi:10.1103/PhysRevLett.113.152003
	[arXiv:1408.0971 [hep-ex]].  
	\bibitem{11} 
	R.~Aaij {\it et al.} [LHCb Collaboration],
	Phys.\ Rev.\ D {\bf 90}, no. 3, 032009 (2014)
	doi:10.1103/PhysRevD.90.032009
	[arXiv:1407.2126 [hep-ex]].
	\bibitem{12}
	R.~Aaij {\it et al.} [LHCb Collaboration],
	Phys.\ Rev.\ D {\bf 87}, no. 11, 112012 (2013)
	Addendum: [Phys.\ Rev.\ D {\bf 89}, no. 1, 019901 (2014)]
	doi:10.1103/PhysRevD.87.112012, 10.1103/PhysRevD.89.019901
	[arXiv:1304.4530 [hep-ex]].
	\bibitem{13}
	R.~Aaij {\it et al.} [LHCb Collaboration],
	JHEP {\bf 1405}, 148 (2014)
	doi:10.1007/JHEP05(2014)148
	[arXiv:1404.0287 [hep-ex]].
	
	\bibitem{14}
	G.~Aad {\it et al.} [ATLAS Collaboration],
	Phys.\ Rev.\ Lett.\  {\bf 113}, no. 21, 212004 (2014)
	doi:10.1103/PhysRevLett.113.212004
	[arXiv:1407.1032 [hep-ex]].
	\bibitem{15}
	R.~Aaij {\it et al.} [LHCb Collaboration],
	Phys.\ Rev.\ Lett.\  {\bf 114}, 132001 (2015)
	doi:10.1103/PhysRevLett.114.132001
	[arXiv:1411.2943 [hep-ex]].
	\bibitem{16}
	R.~Aaij {\it et al.} [LHCb Collaboration],
	Phys.\ Rev.\ D {\bf 92}, no. 7, 072007 (2015)
	doi:10.1103/PhysRevD.92.072007
	[arXiv:1507.03516 [hep-ex]].
	\bibitem{17}
	R.~Aaij {\it et al.} [LHCb Collaboration],
	Phys.\ Rev.\ D {\bf 95}, no. 3, 032005 (2017)
	doi:10.1103/PhysRevD.95.032005
	[arXiv:1612.07421 [hep-ex]].
	
	
	

	\bibitem{18}
	T.~A.~Aaltonen {\it et al.} [CDF Collaboration],
	Phys.\ Rev.\ D {\bf 93} (2016) no.5,  052001
	doi:10.1103/PhysRevD.93.052001
	[arXiv:1601.03819 [hep-ex]].
	\bibitem{19}  
	C.~H.~Chang, C.~F.~Qiao, J.~X.~Wang and X.~G.~Wu,
	Phys.\ Rev.\ D {\bf 72}, 114009 (2005)
	doi:10.1103/PhysRevD.72.114009
	[hep-ph/0509040].
	\bibitem{20}
	H.~Y.~Bi, R.~Y.~Zhang, H.~Y.~Han, Y.~Jiang and X.~G.~Wu,
	Phys.\ Rev.\ D {\bf 95}, no. 3, 034019 (2017)
	doi:10.1103/PhysRevD.95.034019
	[arXiv:1612.07990 [hep-ph]].  
	\bibitem{21} 
	G.~Chen, C.~H.~Chang and X.~G.~Wu,
	Phys.\ Rev.\ D {\bf 97}, no. 11, 114022 (2018)
	doi:10.1103/PhysRevD.97.114022
	[arXiv:1803.11447 [hep-ph]].
	
	\bibitem{22} 
	J.~L.~Abelleira Fernandez {\it et al.} [LHeC Study Group],
	J.\ Phys.\ G {\bf 39}, 075001 (2012)
		doi:10.1088/0954-3899/39/7/075001
		[arXiv:1206.2913 [physics.acc-ph]]. 
	\bibitem{23} 
	A.~Djouadi {\it et al.} [ILC Collaboration],
	arXiv:0709.1893 [hep-ph].
	
	\bibitem{24} 
	J.~Erler, S.~Heinemeyer, W.~Hollik, G.~Weiglein and P.~M.~Zerwas,
	Phys.\ Lett.\ B {\bf 486}, 125 (2000)
	doi:10.1016/S0370-2693(00)00749-8
	[hep-ph/0005024].
	
	\bibitem{25}
	X.~C.~Zheng, C.~H.~Chang, T.~F.~Feng and Z.~Pan,
	Sci.\ China Phys.\ Mech.\ Astron.\  {\bf 61}, no. 3, 031012 (2018)
	doi:10.1007/s11433-017-9121-3
	[arXiv:1701.04561 [hep-ph]].

	\bibitem{26} 
	R.~Aaij {\it et al.} [LHCb Collaboration],
	JHEP {\bf 1801}, 138 (2018)
	doi:10.1007/JHEP01(2018)138
	[arXiv:1712.04094 [hep-ex]].
	
	\bibitem{27}
	R.~J.~Dowdall, C.~T.~H.~Davies, T.~C.~Hammant and R.~R.~Horgan,
	Phys.\ Rev.\ D {\bf 86}, 094510 (2012)
	doi:10.1103/PhysRevD.86.094510
	[arXiv:1207.5149 [hep-lat]].
	
	\bibitem{28} 
	N.~Mathur, M.~Padmanath and S.~Mondal,
	Phys.\ Rev.\ Lett.\  {\bf 121}, no. 20, 202002 (2018)
	doi:10.1103/PhysRevLett.121.202002
	[arXiv:1806.04151 [hep-lat]].
	
	
	
	\bibitem{29} 
	B.~Grinstein and J.~Martin Camalich,
	Phys.\ Rev.\ Lett.\  {\bf 116}, no. 14, 141801 (2016)
	doi:10.1103/PhysRevLett.116.141801
	[arXiv:1509.05049 [hep-ph]].
	\bibitem{30} 
	D.~Kumar, J.~Saini, S.~Gangal and S.~B.~Das,
	Phys.\ Rev.\ D {\bf 97}, no. 3, 035007 (2018)
	doi:10.1103/PhysRevD.97.035007
	[arXiv:1711.01989 [hep-ph]].
	
	\bibitem{31} 
	S.~Sahoo, R.~Mohanta and A.~K.~Giri,
	Phys.\ Rev.\ D {\bf 95}, no. 3, 035027 (2017)
	doi:10.1103/PhysRevD.95.035027
	[arXiv:1609.04367 [hep-ph]].
	
	\bibitem{32}
	G.~Z.~Xu, Y.~Qiu, C.~P.~Shen and Y.~J.~Zhang,
	Eur.\ Phys.\ J.\ C {\bf 76} (2016) no.11,  583
	doi:10.1140/epjc/s10052-016-4423-z
	[arXiv:1601.03386 [hep-ph]].
	
	\bibitem{33}
	S.~Kumbhakar and J.~Saini,
	arXiv:1807.04055 [hep-ph].
	
	\bibitem{34}
	G.~Bonvicini {\it et al.} [CLEO Collaboration],
	Phys.\ Rev.\ Lett.\  {\bf 96}, 022002 (2006)
	doi:10.1103/PhysRevLett.96.022002
	[hep-ex/0510034].
	
	\bibitem{35}
	R.~Louvot {\it et al.} [Belle Collaboration],
	Phys.\ Rev.\ Lett.\  {\bf 102}, 021801 (2009)
	doi:10.1103/PhysRevLett.102.021801
	[arXiv:0809.2526 [hep-ex]].
	
\bibitem{36} 
C.~H.~Chang and Y.~Q.~Chen,
Phys.\ Rev.\ D {\bf 49}, 3399 (1994).
doi:10.1103/PhysRevD.49.3399

\bibitem{37} 
M.~A.~Ivanov, J.~G.~Korner and O.~N.~Pakhomova,
Phys.\ Lett.\ B {\bf 555}, 189 (2003)
doi:10.1016/S0370-2693(03)00052-2
[hep-ph/0212291].

\bibitem{38} 
E.~Hernandez, J.~Nieves and J.~M.~Verde-Velasco,
Phys.\ Rev.\ D {\bf 74}, 074008 (2006)
doi:10.1103/PhysRevD.74.074008
[hep-ph/0607150].

\bibitem{39} 
W.~Wang, Y.~L.~Shen and C.~D.~Lu,
Eur.\ Phys.\ J.\ C {\bf 51}, 841 (2007)
doi:10.1140/epjc/s10052-007-0334-3
[arXiv:0704.2493 [hep-ph]].

\bibitem{40} 
S.~Descotes-Genon, J.~He, E.~Kou and P.~Robbe,
Phys.\ Rev.\ D {\bf 80}, 114031 (2009)
doi:10.1103/PhysRevD.80.114031
[arXiv:0907.2256 [hep-ph]]; and references therein.

\bibitem{41} 
D.~Ebert, R.~N.~Faustov and V.~O.~Galkin,
Eur.\ Phys.\ J.\ C {\bf 32}, 29 (2003)
doi:10.1140/epjc/s2003-01347-5
[hep-ph/0308149].

\bibitem{42} 
D.~Ebert, R.~N.~Faustov and V.~O.~Galkin,
Phys.\ Rev.\ D {\bf 68}, 094020 (2003)
doi:10.1103/PhysRevD.68.094020
[hep-ph/0306306].

\bibitem{43} 
V.~V.~Kiselev, O.~N.~Pakhomova and V.~A.~Saleev,
J.\ Phys.\ G {\bf 28}, 595 (2002)
doi:10.1088/0954-3899/28/4/301
[hep-ph/0110180].

\bibitem{44}
	J.~Sun, Y.~Yang, Q.~Chang and G.~Lu,
	Phys.\ Rev.\ D {\bf 89} (2014) no.11,  114019
	doi:10.1103/PhysRevD.89.114019
	[arXiv:1406.4925 [hep-ph]]; and references therein.
	
\bibitem{45} 
	Z.~T.~Zou, Y.~Li and X.~Liu,
	Phys.\ Rev.\ D {\bf 97}, no. 5, 053005 (2018)
	doi:10.1103/PhysRevD.97.053005
	[arXiv:1712.02239 [hep-ph]].
	
	\bibitem{46}
	S.~Kar, P.~C.~Dash, M.~Priyadarsini, S.~Naimuddin and N.~Barik,
	Phys.\ Rev.\ D {\bf 88} (2013) no.9,  094014.
	doi:10.1103/PhysRevD.88.094014

\bibitem{47} 
R. Dhir, N. Sharma and R.C. Verma,
J. Phys. G. \textbf{35}, 085002 (2008) doi: 10.1088/0954-3899/35/8/085002; and the unpublished work on CKM-suppressed modes.

\bibitem{48} 
R.~Dhir and R.~C.~Verma,
Phys.\ Rev.\ D {\bf 79}, 034004 (2009)
doi:10.1103/PhysRevD.79.034004
[arXiv:0810.4284 [hep-ph]]; and the unpublished work on CKM-suppressed modes.


\bibitem{49}
J.~Sun, Y.~Yang, N.~Wang, Q.~Chang and G.~Lu,
Phys.\ Rev.\ D {\bf 95}, no. 7, 074032 (2017)
doi:10.1103/PhysRevD.95.074032
[arXiv:1705.09477 [hep-ph]].

\bibitem{50}
Q.~Chang, L.~L.~Chen and S.~Xu,
J.\ Phys.\ G {\bf 45}, no. 7, 075005 (2018)
doi:10.1088/1361-6471/aac732
[arXiv:1806.02076 [hep-ph]].

\bibitem{51}
J.~Sun, Y.~Yang, N.~Wang, J.~Huang and Q.~Chang,
Phys.\ Rev.\ D {\bf 95}, no. 3, 036024 (2017)
doi:10.1103/PhysRevD.95.036024
[arXiv:1703.00155 [hep-ph]].

\bibitem{52} 
Q.~Chang, P.~P.~Li, X.~H.~Hu and L.~Han,
Int.\ J.\ Mod.\ Phys.\ A {\bf 30}, no. 27, 1550162 (2015)
doi:10.1142/S0217751X15501626
[arXiv:1605.01630 [hep-ph]].

\bibitem{53}  
T.~Wang, Y.~Jiang, T.~Zhou, X.~Z.~Tan and G.~L.~Wang,
J.\ Phys.\ G {\bf 45}, no. 11, 115001 (2018)
doi:10.1088/1361-6471/aae14a
[arXiv:1804.06545 [hep-ph]].

\bibitem{54} 
R.~Dhir, R.~C.~Verma and A.~C. Sharma
Adv.\ High Energy Phys.\  {\bf 2013}, 706543 (2013)
doi:10.1155/2013/706543
[arXiv:0903.1201 [hep-ph]].

\bibitem{55} 
M.~Wirbel, B.~Stech and M.~Bauer,
Z.\ Phys.\ C {\bf 29}, 637 (1985) doi: 10.1007/BF01560299.

\bibitem{56}
M.~Bauer, B.~Stech and M.~Wirbel, 
Z.\ Phys. \ C {\bf 34}, 103 (1987) doi: 10.1007/BF01561122.
\bibitem{57}  
M.~Wirbel, Prog. Part. Nucl. Phys. \textbf{21}, 33(1988) doi: 10.1016/0146-6410(88)90031-2.

\bibitem{58}
G.~Buchalla, A.~J.~Buras and M.~E.~Lautenbacher,
Rev.\ Mod.\ Phys.\  {\bf 68}, 1125 (1996)
doi:10.1103/RevModPhys.68.1125
[hep-ph/9512380]. doi:10.1103/RevModPhys.68.1125.

\bibitem{59}  
M.~Neubert, V.~Riekert, Q. P.~Xu, and B.~Stech,
"Exclusive weak decays of \textit{B}-mesons" in "Heavy Flavors", ed.
A.J.~Buras and H.~Linder (Singapore: World Scientific) {\bf 28} (1992). 
\bibitem{60}  
T.~E.~Browder and K.~Honscheid,	
Prog.\ Part.\ Nucl.\ Phys.\  {\bf 35}, 81 (1995)
doi:10.1016/0146-6410(95)00042-H
[hep-ph/9503414].

\bibitem{61} 
A.~Ali, J.G.~Korner, G.~Kramer and J.~Willordt, 
Z. Phys. C \textbf{1}, 269 (1979) doi: 10.1007/BF01440227.
\bibitem{62} 
A.~Ali, J.G.~Korner, G.~Kramer and J.~Willordt,
Z. Phys. C \textbf{2}, 33 (1979) doi: 10.1007/BF01546234.
\bibitem{63} 
G. Kramer and W.F. Palmer,
Phys. Rev. D \textbf{45}, 193 (1992) doi: 10.1103/PhysRevD.45.193.

\bibitem{64} 
H.~Y.~Cheng,
Phys.\ Rev.\ D {\bf 67}, 094007 (2003)
doi:10.1103/PhysRevD.67.094007
[hep-ph/0301198].

\bibitem{65}
D.H. Perkins,
(Cambridge University Press 2000) 4th ed., p.127. 

\bibitem{66} 
M.~Tanabashi {\it et al.} [Particle Data Group],
Phys.\ Rev.\ D {\bf 98}, no. 3, 030001 (2018); and 2019 update.
doi:10.1103/PhysRevD.98.030001

\bibitem{67} 
A.~Deur, S.~J.~Brodsky and G.~F.~de Teramond,
Prog.\ Part.\ Nucl.\ Phys.\  {\bf 90}, 1 (2016)
doi:10.1016/j.ppnp.2016.04.003
[arXiv:1604.08082 [hep-ph]]; and references therein.

\bibitem{68} 
G.~Ganbold,
Particles {\bf 2}, no. 2, 180 (2019)
doi:10.3390/particles2020013; ibid. arXiv:1712.05531 [hep-ph]; and references therein.

\bibitem{69} 
E.~P.~Biernat, F.~Gross, M.~T.~Peña, A.~Stadler and S.~Leitão,
Phys.\ Rev.\ D {\bf 98}, no. 11, 114033 (2018)
doi:10.1103/PhysRevD.98.114033
[arXiv:1811.01003 [hep-ph]].

\bibitem{70} 
R.~J.~Crewther and L.~C.~Tunstall,
Phys.\ Rev.\ D {\bf 91}, no. 3, 034016 (2015)
doi:10.1103/PhysRevD.91.034016
[arXiv:1312.3319 [hep-ph]].

\bibitem{71} 
G.~Ganbold,
Phys.\ Rev.\ D {\bf 81}, 094008 (2010)
doi:10.1103/PhysRevD.81.094008
[arXiv:1004.5280 [hep-ph]].

\bibitem{72} 
  K.~G.~Chetyrkin, J.~H.~Kuhn and M.~Steinhauser,
  Comput.\ Phys.\ Commun.\  {\bf 133}, 43 (2000)
  doi:10.1016/S0010-4655(00)00155-7
  [hep-ph/0004189].
 
 \bibitem{73} 
  C.~Quigg and J.~L.~Rosner,
  Phys.\ Lett.\  {\bf 71B}, 153 (1977).
  doi:10.1016/0370-2693(77)90765-1

\bibitem{74} 
A.~Martin,
Phys.\ Lett.\  {\bf 93B}, 338 (1980).
doi:10.1016/0370-2693(80)90527-4

\bibitem{75} 
W.~Buchmuller and S.~H.~H.~Tye,
Phys.\ Rev.\ D {\bf 24}, 132 (1981).
doi:10.1103/PhysRevD.24.132

\bibitem{76}
E.~J.~Eichten and C.~Quigg,
Phys.\ Rev.\ D {\bf 49} (1994) 5845
doi:10.1103/PhysRevD.49.5845
[hep-ph/9402210].

\bibitem{77}
M.~Karliner and J.~L.~Rosner,
Phys.\ Rev.\ D {\bf 90} (2014) no.9,  094007
doi:10.1103/PhysRevD.90.094007
[arXiv:1408.5877 [hep-ph]].

\bibitem{78}
M.~Karliner and J.~L.~Rosner,
Phys.\ Rev.\ Lett.\  {\bf 119} (2017) no.20,  202001
doi:10.1103/PhysRevLett.119.202001
[arXiv:1707.07666 [hep-ph]].

\bibitem{79}
E.~J.~Eichten and C.~Quigg,
Phys.\ Rev.\ D {\bf 99} (2019) no.5,  054025
doi:10.1103/PhysRevD.99.054025
[arXiv:1902.09735 [hep-ph]].

\bibitem{80} 
N.~Brambilla, A.~Pineda, J.~Soto and A.~Vairo,
Rev.\ Mod.\ Phys.\  {\bf 77}, 1423 (2005)
doi:10.1103/RevModPhys.77.1423
[hep-ph/0410047].

\bibitem{81} 
T. Mannel, Effective Field Theories for Heavy Quarks - 2017, Les Houches Summer School in Theoretical Physics.

\bibitem{82} 
T.~Mannel and S.~Wolf,
hep-ph/9701324.



\bibitem{83} 
G.~P.~Lepage, L.~Magnea, C.~Nakhleh, U.~Magnea and K.~Hornbostel,
Phys.\ Rev.\ D {\bf 46}, 4052 (1992)
doi:10.1103/PhysRevD.46.4052
[hep-lat/9205007].

\bibitem{84}
Y.~Yang, J.~Sun, Y.~Guo, Q.~Li, J.~Huang and Q.~Chang,
Phys.\ Lett.\ B {\bf 751} (2015) 171
doi:10.1016/j.physletb.2015.10.018
[arXiv:1701.04593 [hep-ph]].

\bibitem{85} 
H.~Y.~Cheng, C.~K.~Chua and C.~W.~Hwang,
Phys.\ Rev.\ D {\bf 69}, 074025 (2004)
doi:10.1103/PhysRevD.69.074025
[hep-ph/0310359].

\bibitem{86}
P.~Gelhausen, A.~Khodjamirian, A.~A.~Pivovarov and D.~Rosenthal,
Phys.\ Rev.\ D {\bf 88}, 014015 (2013)
Erratum: [Phys.\ Rev.\ D {\bf 89}, 099901 (2014)]
Erratum: [Phys.\ Rev.\ D {\bf 91}, 099901 (2015)]
doi:10.1103/PhysRevD.88.014015, 10.1103/PhysRevD.91.099901, 10.1103/PhysRevD.89.099901
[arXiv:1305.5432 [hep-ph]].

\bibitem{87}
C.~McNeile, C.~T.~H.~Davies, E.~Follana, K.~Hornbostel and G.~P.~Lepage,
Phys.\ Rev.\ D {\bf 86}, 074503 (2012)
doi:10.1103/PhysRevD.86.074503
[arXiv:1207.0994 [hep-lat]].


	\bibitem{88}
	V. SŠimonis,
	Eur.\ Phys.\ J.\ A {\bf 52}, no. 4, 90 (2016)
	doi:10.1140/epja/i2016-16090-5
	[arXiv:1604.05894 [hep-ph]].
	
	\bibitem{89}
	D.~Ebert, R.~N.~Faustov and V.~O.~Galkin,
	Phys.\ Lett.\ B {\bf 537}, 241 (2002)
	doi:10.1016/S0370-2693(02)01939-1
	[hep-ph/0204089].
	

	
	
	

	

	
	

	
	
\end{thebibliography}
\end{document}